\documentclass[a4paper,twocolumn,superscriptaddress,11pt,accepted=2021-04-22]{quantumarticle}
\pdfoutput=1

\usepackage[numbers,sort&compress]{natbib}
\usepackage[utf8]{inputenc}
\usepackage[english]{babel}
\usepackage[T1]{fontenc}
\usepackage{amsmath}
\usepackage{color}
\usepackage{graphicx}
\usepackage{dcolumn}
\usepackage{array}
\usepackage{float}
\usepackage{longtable}
\usepackage{mathrsfs}
\usepackage{txfonts}
\usepackage{wasysym}
\usepackage[T1]{fontenc}
\usepackage{amssymb}
\usepackage[usenames,dvipsnames]{xcolor}
\usepackage{amsmath}
\usepackage{amstext}
\usepackage{latexsym}
\usepackage{hyperref}
\usepackage{bbold}

\usepackage{amsbsy} 

\def\la{\langle}
\def\ra{\rangle}

\def\tr{{\rm Tr}}

\def\dg{^\dagger}
\newcommand{\beq}{\begin{equation}}
\newcommand{\eeq}{\end{equation}}
\newcommand{\beqa}{\begin{eqnarray}}
\newcommand{\eeqa}{\end{eqnarray}}

\newcommand{\ket}[1] {\vert #1 \rangle}
\newcommand{\bra}[1] {\langle #1 |}
\newcommand{\braket}[2] {\langle #1 | #2 \rangle}
\newcommand{\Tr}{\mathrm{Tr}}

\mathchardef\myomega="0121

\newcommand{\Tdil}{S_{r,0}}
\newcommand{\Hsq}{H_{\textsc{gho}}}

\graphicspath{{../images/}}

\begin{document}

\title{Shortcuts to Squeezed Thermal States}

\author{L\'eonce Dupays  \href{https://orcid.org/0000-0002-3450-1861}{\includegraphics[scale=0.05]{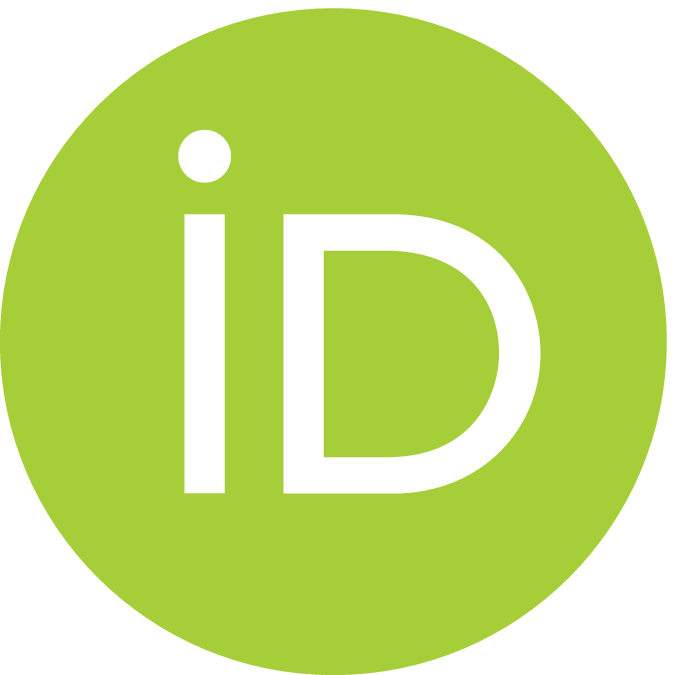}}}
\affiliation{Donostia International Physics Center,  E-20018 San Sebasti\'an, Spain } 
\affiliation{Department of Physics and Materials Science, University of Luxembourg, L-1511 Luxembourg, G.D. Luxembourg}

\author{Aur\'elia Chenu \href{https://orcid.org/0000-0002-4461-8289}{\includegraphics[scale=0.05]{orcidid.eps}}}
\email{aurelia.chenu@uni.lu}
\affiliation{Donostia International Physics Center,  E-20018 San Sebasti\'an, Spain } 
\affiliation{Department of Physics and Materials Science, University of Luxembourg, L-1511 Luxembourg, G.D. Luxembourg}
\affiliation{Ikerbasque, Basque Foundation for Science, E-48013 Bilbao, Spain}

\maketitle


\begin{abstract}
  Squeezed state in harmonic systems can be generated through a variety of techniques, including varying the oscillator frequency or using nonlinear two-photon Raman interaction. We focus on these two techniques to drive an initial thermal state into a final squeezed thermal state  with controlled squeezing parameters---amplitude and phase---in arbitrary time. The protocols are designed through reverse engineering for both unitary and open dynamics. Control of the dissipation is achieved using stochastic processes, readily implementable via, e.g., continuous quantum measurements. Importantly, this allows  controlling the state entropy and can be used for fast thermalization. The developed protocols are thus suited to generate squeezed thermal states at controlled temperature in arbitrary time.
  \end{abstract}

\section{Introduction}

Squeezing is a paradigmatic quantum effect that allows reducing fluctuations of one variable beneath the standard quantum limit. This is achieved at the expenses of increasing the variance of the conjugated variable, such that Heisenberg uncertainty principle still holds true for the product of the variances. 
Squeezed states have kept their promise in improving measurement accuracy beyond quantum noise \cite{Abadie2011a, wolfgramm2010} and have become central in quantum optics \cite{polzik2008} through demonstrated applications in quantum metrology and sensing \cite{kimble1987, polzik1992}. 
Advanced techniques to generate squeezed light \cite{vahlbruch2008, takeno2007, mccormick2007} facilitated the detection of gravitational wave \cite{caves1980, goda2008, Tse2019}. Theoretical works have proposed applications in quantum information, where coupling a qubit to a squeezed reservoir allows  erasing information below the Landauer's limit \cite{klaers2019}. In the context of quantum thermodynamics, the proposed theories of coupling the working medium of a nanoscale heat engine to a squeezed reservoir to generate work beyond the Carnot's limit \cite{huang2012,rossnagel2014, correa2014, manzano2016, niedenzu2016, niedenzu2018} have been experimentally demonstrated  using a vibrating nano-beam driven by squeezed electronic noise     \cite{klaers2017}.

\begin{figure}[b!]
\includegraphics[width=1\columnwidth]{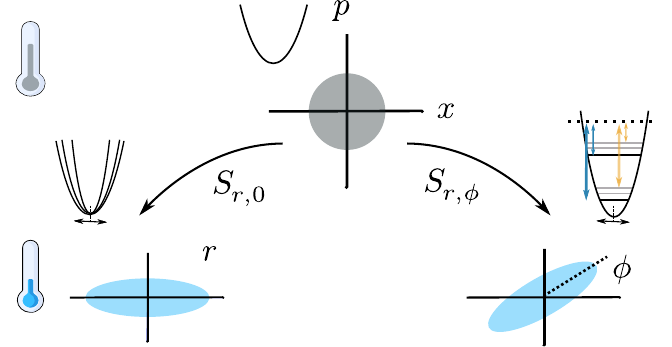}
\captionof{figure}{Schematic representation of the control processes considered in this work. Starting from an initial thermal state with isotropic density in phase-space (top), we design dynamical protocols to generate a squeezed thermal state (bottom) at controlled temperature in arbitrary time using two different experimental implementations: (left) in a harmonic oscillator with controlled frequency, or (right) using two-photon Raman interaction. Thermalization is achieved by  engineering a dissipator in position. \label{fig:scheme}}
\end{figure}

Renewed interest in squeezing has come with progress in quantum optomechanics \cite{Rashid2016,kippenberg2008, meystre2013, aspelmeyer2014,asjad2014}. In a simple parametric interaction where the spring constant of a mechanical oscillator is controlled with radiation pressure forces, the emergence of mechanical instabilities prevents reducing fluctuations to at most 50\% (the 3 dB limit) in the steady state \cite{milburn1981, szorkovszky2011}. Schemes to generate squeezing below this limit have been developed by combining parametric driving and weak measurements \cite{szorkovszky2011, szorkovszky2013}. Continuous measurements, where measuring one variable---e.g. position---precisely reduces its associated variance, present an intuitive technique that has been broadly explored \cite{caves1981, braginsky1980, ruskov2005, clerk2008, genoni2013, genoni2015, brunelli2019, meng2019, bennett2017, huang2020}.  Alternatively, a simple but powerful scheme that lifts the requirement of  explicit measurement or feedback has been put forward using a dissipative mechanism where the driven cavity  acts as an engineering reservoir \cite{kronwald2013}. This theoretical protocol shows the similarities between processes relying on coherent feedback or reservoir engineering  \cite{abah2014, rossnagel2014}, and has been experimentally demonstrated in \cite{wollman2015, pirkkalainen2015}.

While much progress has been achieved for increasing the squeezing parameter, 
the protocols have  so far been restricted to unitary dynamics and do not allow for the control of entropy. 
We here lift this limitation 
and provide protocols to generate squeezed thermal states at controlled temperature in a fixed time. Controlling the temperature of squeezed thermal states is all the more relevant since the variance of such states not only depends on the  squeezing amplitude but also on the average thermal phonon number \cite{kim1989a, kral1990,walls1983}, as further discussed below.

Specifically, we focus on two different methods, already known as useful to generate squeezing, that we extend to open setups: (i) squeezing from non-adiabatic driving of the oscillator frequency---how is the  trap control frequency modified by the dissipative dynamics; (ii) squeezing through the use of two-photon Raman interaction. 
In both cases, knowledge of the analytical dynamics allows finding the control processes through reverse engineering. Thermalization is achieved through the use of white noise with controlled amplitude, that generates the open dynamics.

The presented protocols aim at the dynamical control of states without relying on adiabatic evolution. In this sense, they fall under the umbrella of Shortcuts to Adiabaticity (STA) \cite{del_Campo_2019,Chen2010}.    In essence, STA provide the control Hamiltonian to generate in a fixed time the state otherwise reached through a reference adiabatic trajectory.
Extension of STA to open quantum systems requires, in addition to the control Hamiltonian, a control dissipator.  This was first explored for Markovian dynamics \cite{Vacanti2014} and a general scheme has recently been put forward for arbitrary dynamics \cite{alipour2020}. Other works have shown how to control the thermalization of a harmonic oscillator \cite{Dann2019a,villazon2019,dupays2020}. Here, we provide protocols combining squeezing and thermalization. Note that squeezed thermal states have been experimentally achieved in, e.g., a massive mechanical harmonic oscillator using sudden quenches \cite{Rashid2016}. Our protocols allow to operate in the quantum regime and avoid creation of excitations thanks to the use STA techniques. Also, they allow us to control the state squeezing parameters as well as its temperature. 
 
The paper is organized as follow: Section \ref{sec:thermalstate} presents the general evolution for a squeezed thermal state with time-dependent parameters. Section \ref{sec:phi0} focuses on the harmonic oscillator setup, clarifying the relations of known STA with squeezing. Section \ref{sec:2photon} presents the results using two-photon Raman interaction that allows controlling the squeezing amplitude, phase, and state temperature.

\section{Squeezed thermal  states \label{sec:thermalstate}}
We consider the thermal state $\sigma_0 = e^{- \beta_0 H_0}/\Tr(e^{- \beta_0 H_0})$ of a harmonic oscillator (HO) with Hamiltonian $H_0 = \hbar \omega_0 (a\dg_0 a_0 + 1/2)$. 
This initial thermal state is driven to a target squeezed thermal state $\sigma_f$ at an arbitrary final time $t_f$ following  
\begin{equation} \label{eq:rho_sq}
 \sigma_{t} = \frac{1}{Z_t} S_{r,\phi} e^{- \frac{\varepsilon_t}{\varepsilon_0}\beta_0 H_0} S_{r,\phi}\dg,
\end{equation} 
where we allow for changes in entropy through the dimensionless time-dependent parameter $\varepsilon_t = \hbar \omega_t \beta_t$ and its initial value $\varepsilon_0 = \hbar \omega_0 \beta_0$. The partition function  $Z_t$ normalizes the state. 
The squeezing operator
\begin{equation}\label{eq:S}
S_{r, \phi} = \exp\left(\frac{r_t}{2} \left(e^{- i \phi_t}  a_0^2 -e^{ i \phi_t}  a^{\dagger 2}_0\right)\right)
\end{equation} 
 is defined from the squeezing parameter $z_t = \frac{r_t}{2} e^{- i\phi_t}$ with time-dependent amplitude $r_t$ and phase $\phi_t$, which defines the correlations between position and momentum. The annihilation operator $a_0$ is defined from 
\begin{equation} \label{eq:at}
a_t = \sqrt{\frac{m \omega_t}{2 \hbar}}  \hat{x} + i \sqrt{\frac{1}{2 \hbar m \omega_t}} \hat{p}. 
\end{equation}
The static properties of state (\ref{eq:rho_sq}) have been thoroughly studied and described in e.g. \cite{LoudonBook}. We focus on its dynamics to design control protocols generating a squeezed thermal state at arbitrary final temperature $\beta_f^{-1}$ and target parameters $\{r_f, \phi_f\}$ at the end of the control protocol, $t_f$. 


 Let us first rewrite (\ref{eq:rho_sq}) as the thermal state of the squeezed harmonic oscillator. Using  $S_{r,\phi} S_{r,\phi}\dg=\mathbbm{1}=S_{r,\phi}\dg S_{r,\phi}$, we have
\begin{equation}\label{eq:rhoSQ}
\sigma_t = \frac{1}{Z_t}e^{-\beta_t \Hsq}
\end{equation}
where $\Hsq =\frac{\omega_t}{\omega_0} S_{r,\phi} H_0 S\dg_{r,\phi}$ and the partition explicitly reads $Z_t=\Tr(e^{- \beta_t \Hsq})$. We split $S_{r,\phi} a\dg_0 a_0 S_{r,\phi}\dg = (S_{r,\phi} a\dg_0  S_{r,\phi}\dg) (S_{r,\phi} a_0 S_{r,\phi}\dg) \equiv A\dg_t A_t$ to define new creation and annihilation operators, explicitly found using the Baker-Campbell-Hausdorf (BCH) formula \cite{MandelWolfBook} 
 , yielding 
\begin{equation}
A_t \equiv S_{r,\phi} a_0  S_{r,\phi}\dg {=} \cosh r_t a_0 {+} e^{i \phi_t} \sinh r_t a\dg_0.  \label{ct} 
\end{equation}
These `$A$' operators are bosonic operators fulfilling $[A_t,A_t\dg]=1$. For each $A\dg_t$ boson created, there is both creation and annihilation of some `$a$' bosons\footnote{Squeezing then appears similar to the physical setup of the independent-boson model \cite{Mahanbook}, which is best dealt with using two different basis for the bosons \cite{chenu2019}.} In this basis, the squeezed harmonic oscillator simply reads  
\begin{eqnarray} \label{eq:Hsq}
\Hsq =\hbar \omega_t (A_t\dg A_t+1/2). 
\end{eqnarray}
%
We proceed to describe the evolution of  its  thermal state.

Under unitary evolution, the state dynamics  $\dot{\sigma}_t = - \frac{i}{\hbar} [\Hsq+H_\textsc{cd},\sigma_t]$  is governed by  
$H_{\textsc{cd}} \equiv  \frac{\hbar}{i } S_{r,\phi} \dot{S}\dg_{r, \phi}$, known as the counter-diabatic (CD) Hamiltonian \cite{muga2010a,delcampo13,Funo17}, 
 which ensures that each eigenstate remains instantaneous eigenstate of the time-dependent Hamiltonian and evolves as $i \hbar \ket{\dot{n}_t} = H_{\textsc{cd}} \ket{n_t}$.  
The explicit form of this Hamiltonian is obtained from the $k$th time-derivatives of $(e^{- i \phi_t}a_0^2 - e^{ i \phi_t}a_0^{\dagger 2})$---see App. \ref{app:SSdot}---yielding 
\begin{eqnarray} \label{H3}
H_{\textsc{cd}} &=&\hbar \frac{\dot{\phi}_t}{2} (A_t\dg A_t +\frac{1}{2}- (a_0\dg a_0+ \frac{1}{2})) \nonumber   \\
&+& i \hbar \frac{\dot{r}_t}{2}(a_0^2e^{- i \phi_t} { -} a^{\dagger 2}_0 e^{i \phi_t}). 
\end{eqnarray}
Let us comment on the possible implementations set by the phase.

 \emph{\underline{Case $\phi_{t}=0$}: (Section \ref{sec:phi0})} 
Squeezing with no final correlation between $\hat{x}$ and $\hat{p}$, i.e. using $S_{r,0}$, reduces the Hamiltonian (\ref{H3}) to
\begin{equation} \label{H30}
H^{(0)}_{\textsc{cd}} = i \hbar \frac{\dot{r}_t}{2}(a_0^2 { -} a^{\dagger 2}_0 ) \\ 
\end{equation}
and the evolution reads $\dot{\sigma}_t = - \frac{i}{\hbar} [\Hsq^{(0)} + H_\textsc{cd}^{(0)},\sigma_t]$. 
We elaborate in Sec. \ref{sec:trapcontrol} how this Hamiltonian can be implemented in a harmonic oscillator by varying the trap frequency and relates to common STA techniques. Indeed, since $-i \hbar (a_0^2 - a_0^{\dagger 2}) = \{\hat{x}, \hat{p}\}$, the counterdiabatic term recovers the known squeezing term  \cite{muga2010a} such that, for a proper choice of $r_t$, it can be recast into the form that, in first quantization, reads  $\Hsq^{(0)}+ H_{\textsc{cd}}^{(0)} =\frac{\hat{p}^2}{2m} + \frac{1}{2} m\omega_t^2 \hat{x}^2-\frac{\dot{r}_t}{2}\{\hat{x}, \hat{p}\}.$
\\

\emph{\underline{Case $\phi_{t}\neq0$}:  (Section \ref{sec:2photon})} 
The generalized HO $ \hbar \frac{\dot{\phi}_t}{2} (A_t\dg A_t +\frac{1}{2})$ commutes with the dynamical state and can be removed from the Hamiltonian. The remaining Hamiltonian in Eq. (\ref{H3}) can be implemented in a rotated frame, as we detail later in Sec. \ref{sec:2photon}.  For now, note that the unitary rotation $U_{\phi}=e^{-i\frac{\phi_t}{2}(a^{\dagger}_{0}a_{0}+\frac{1}{2})}$ leads to the state $\tilde{\sigma}_{t}=U_{\phi}\sigma_{t}U^{\dagger}_{\phi}$ evolving as\footnote{The rotation $R(\theta)=e^{-i\frac{\theta}{2}a^{\dagger}_{0}a_{0}}$ verifies the property $R(\theta)S_{r,\phi}R^{\dagger}(\theta)=S_{r,\phi-\theta}$. It is thus straightforward to cancel the phase dependence with a rotation $U_{\phi}=R(\phi_{t})e^{-i \frac{\phi_{t}}{ 4}}$.}, 
\begin{equation}
\frac{d\tilde{\sigma}_{t}}{dt}=\frac{1}{i\hbar}[ i \hbar \frac{\dot{r}_t}{2}(a_0^2{ -} a^{\dagger 2}_0 ),\tilde{\sigma}_{t}].\label{eq:dilatation}
\end{equation}
This unitary evolution is dictated by the two-photon Raman Hamiltonian. We further detail how this allows implementing (\ref{H3}) provided that $\dot{\phi}_t=-2\omega_{0}$.  Importantly, the phase linearly depends on the process time. So with this setup, a process of fixed time $t_f$ generates a fixed squeezing phase, $\phi_{f}=-2\omega_{0}t_{f}$. 
We show in Sec. \ref{sec:2photon} how to lift this constraint and generate a squeezed state with arbitrary position-momemtum correlation in arbitrary time. 

\section{Fast Squeezing and Thermalization through trap and dephasing control \label{sec:phi0}}
On  one hand, it is quite established that squeezing can be achieved with a change of the trap frequency, as proposed in trapped ions  already decades ago \cite{Lo_1990,heinzen1990,lau2012} and demonstrated experimentally  \cite{wineland1998}. 
On  the other hand, creating a thermal state from the thermal state of a different system requires 
rearrangement of the initial distribution of eigenstates so as to match the Gibbs distribution of the final system. STA protocols generate, in a finite time, the adiabatic evolution of a reference  Hamiltonian  \cite{Chen2010}, thus preserving the initial eigenvalue distribution. 
Although these two results are well established, the connection between squeezing and STA on a HO seems to not always be made. We explicit it here and then consider dissipative dynamics to extend the technique to generate squeezed thermal state  at arbitrary temperature.  To do so, this Section focuses on generating squeezed thermal state with no correlation between $\hat{x}$ and $\hat{p}$, i.e. $\phi=0$.

\subsection{Squeezing through trap control \label{sec:trapcontrol}}
 We first consider unitary, isentropic dynamics, i.e. $\varepsilon_t = \varepsilon_0$ constant. In the case $\phi_t =0$, the squeezed thermal state directly maps to the instantaneous thermal state  of a HO with time-dependent frequency $\omega_t$, 
\begin{equation}\label{sigmat}
\sigma_t(\phi=0) = S_{r,0} \sigma_0 S_{r,0}\dg = \frac{e^{-\beta_t H_t}}{\tr\left(e^{-\beta_t H_t}\right)}.
\end{equation}
Indeed, for $\phi_t=0$, the $A_t$ operator (\ref{ct})  becomes $a_t=\Tdil \,a_0\,\Tdil\dg = a_0 \cosh r_t +a\dg_0 \sinh r_t$. It evolves as $\dot{a}_t = \dot{r}_t a\dg_t$ which maps, for $r_t = \ln \sqrt{\omega_t / \omega_0}$, to the annihilation operator factorizing the time-dependent HO, $\Hsq^{(0)} = H_t = \hbar \omega_t (a\dg_t a_t + \frac{1}{2})$.

Note that the operation $S_{r,0}$ is also known as a dilatation  \cite{lohe2009}, $T_w = \exp\big(-\frac{i\log(w)}{2\hbar}(\hat{x}\hat{p}+\hat{p}\hat{x})\big)$ with  $w \equiv \sqrt{\omega_{0} / \omega_{t}}$, that  transforms position and momentum  as $\Tdil f(\hat{x}) \Tdil\dg = f(\hat{x}/w)$ and $\Tdil f(\hat{p}) \Tdil\dg = f(w \hat{p})$, respectively. 
The time-dependent HO itself is thus equivalently a squeezed or dilated HO
\begin{equation}
H_t =\frac{\omega_t}{\omega_0}S_{r_t, 0} H_0 S_{r_t, 0}\dg = T_w \frac{H_0}{w^2}T_w^\dag.
\end{equation}
The dynamics directly follows from (\ref{H30}) as $\dot{\sigma}_t = -\frac{i}{\hbar}[H_t+H_{\textsc{cd}}^{(0)}, \sigma_t]$. 

For the purpose of experimental implementation, let us consider the state $\varrho_t = U_{\Omega_0}  \sigma_t U_{\Omega_0}\dg$ in a frame rotated by the unitary  $U_{\Omega_0} = \exp\big({i\frac{\Omega_{0} m }{2 \hbar} \hat{x}^2} \big)= \exp\big( i \frac{\Omega_0}{4\omega_t}\big(a_{t} + a\dg_{t}\big)^2\big)$, where $\Omega_0$ so far is an arbitrary, time-dependent frequency. The evolution of this density matrix  $\dot{\varrho}_t = - \frac{i}{\hbar} [H_{c}, \varrho_t]$  is governed 
 by the control Hamiltonian $H_{c} = U_{\Omega_0}(H_t+H_{\textsc{cd}}^{(0)})U_{\Omega_0}\dg + i \hbar \dot{U}_{\Omega_0} U_{\Omega_0}\dg$. Using the fact that $U_{\Omega_0} a_t U_{\Omega_0}\dg = a_t - i \frac{\Omega_0}{2 \omega_t}(a_t+a_t\dg)$, this Hamiltonian reads\footnote{We used $ a_t^2 - a^{\dagger 2}_t=(\cosh(r_t)a_0+\sinh(r_t)a^{\dagger}_0)^{2}-(\cosh(r_t)a^{\dagger}_{0}+\sinh(r_t)a_{0})^{2}=a_0^2 - a^{\dagger 2}_0 $. }
\begin{eqnarray}\label{eq1}  
H_{c} &=&\hbar \omega_t (a_t\dg a_t + \frac{1}{2}) \nonumber \\
&&+  \frac{\hbar}{4 \omega_t} \left( \Omega_0^2+ \Omega_0 \frac{\dot{\omega}_t}{\omega_t}- \dot{\Omega}_0\right)(a_t+a_t\dg)^2 \nonumber \\
&&+ i \hbar \left(\frac{\Omega_0}{2} + \frac{\dot{\omega}_t}{4 \omega_t}\right)(a_t^2 - a_t^{\dagger 2}).  
\end{eqnarray}

Taking $\Omega_0 = -\dot{\omega}_t/ (2 \omega_t)=-\dot{r}_t= \dot{w}/w$ removes the correlations in position and momentum. Under this condition, the control Hamiltonian $H_{c}=\frac{\hat{p}^{2}}{2m}+\frac{1}{2}m\omega^{2}_{c}\hat{x}^{2}$ is a HO with time-dependent control  frequency (\ref{weff})
\begin{eqnarray}\label{weff}
\omega_{c}^2 &=&\omega^{2}_{t}-\Omega^{2}_{0}-\dot{\Omega}_{0}\nonumber\\
 &=&\omega_t^2 -\frac{3}{4} \frac{\dot{\omega}_t^2}{\omega_t^2}+\frac{1}{2}\left( \frac{\ddot{\omega}_t}{\omega_t}\right).
\end{eqnarray} 
 So for a given reference trajectory with frequency $\omega_{t}$, $\omega_{c}$ is the frequency to be implemented to do the shortcut. This control frequency can be recovered inserting $w$ in the Ermakov equation \cite{ermakov1880},
 $\ddot{w} + \omega_{\rm eff}^2 w = \omega_0^2 / w^3$, and corresponds to known results from local counterdiabatic driving \cite{delcampo13,DeffnerJarzynskiAdC14}. 
The state implemented in the lab evolves as
$\varrho_t=U_{ \Omega_{0}} S_{r,0}  \sigma_0 S_{r,0}\dg U^{\dagger}_{ \Omega_{0}}$ and corresponds to the target squeezed state at the end of the process only---at which time $U_{\Omega_{0}}$  becomes the identity and $\varrho_f= \sigma_f (\phi=0)$.

\subsection{Extending the range of accessible squeezed states with a control dissipator}
The  thermal state (\ref{sigmat})  is diagonal in the instantaneous Fock state basis and reads $\sigma_t = \sum_n p_{n,t}\ket{n_t}\bra{n_t}$, with  $p_{n,t} = e^{-\varepsilon_t n} (1 - e^{-\varepsilon_t})$.
  Its von Neumann entropy  $- \Tr(\sigma_t \ln \sigma_t) = \varepsilon_t/(e^{\varepsilon_t}-1) - \ln(1 - e^{- \varepsilon_t})$ is conserved during any unitary evolution. 
The product  $\beta\omega$, that characterizes the eigenvalue distribution of the thermal state, is  constant under unitary evolution. We refer to `cooling' for processes decreasing the von Neumann entropy \cite{ketterle1992}, which contrast with phase-space preserving processes---with constant entropy. To extend the range of accessible states, we allow for changes in temperature and entropy  during the dynamics, taking $\varepsilon_t$  time dependent.

\begin{figure}
\begin{center}
\includegraphics[width=1\columnwidth]{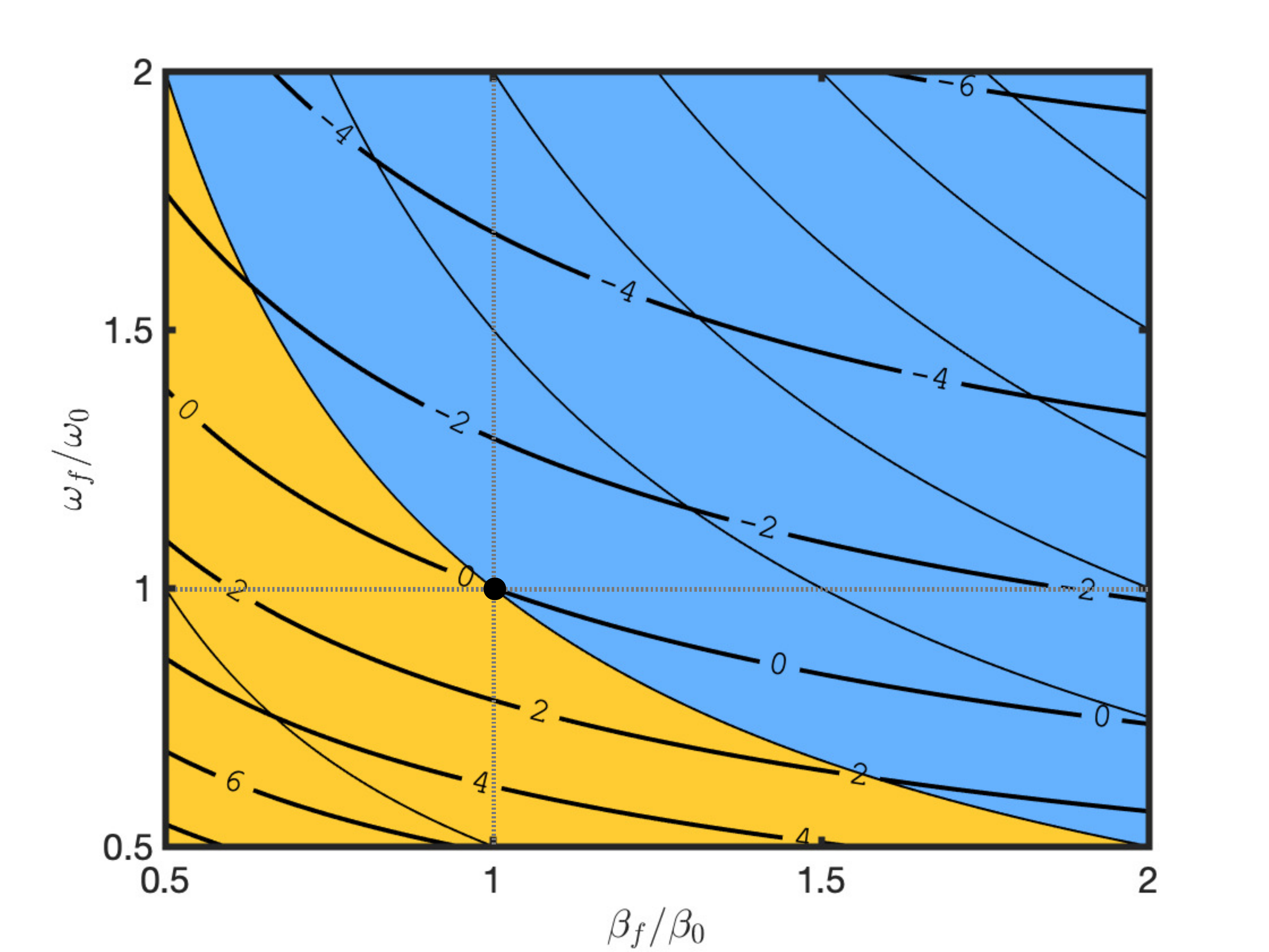}
\caption{\textbf{Map of accessible variances:} The bold lines correspond to constant relative variances, with values $10\log_{10}(\frac{ \Delta x_{f} }{ \Delta x_{0}  })$ given in dB. 
The black point represents the identity process. The light contour lines correspond to isentropic processes (constant $\beta\omega$). Unitary dynamics are along these isentropic lines and restrict the target state to $\beta_{f}\omega_{f}=\beta_{0}\omega_{0}$, thus also restricting the accessible variances for given initial conditions. By contrast, open dynamics and engineered dissipation extend the variances at reach to the full map. Blue background correspond to cooling processes, orange is for heating. \label{variancemap} }
\end{center}
\end{figure}

Another motivation to design open protocols for squeezing is to enhance the variance. Indeed, the variance in position of a squeezed thermal state depends on both the average phonon number $\bar{n}= \frac{1}{e^{\varepsilon_{t}}-1}$ and the squeezing parameter $r_{t}$ \cite{kim1989a}. Specifically, using the operator $\hat{x}=\sqrt{\frac{\hbar}{m\omega_{0}}}(a_{0}+a^{\dagger}_{0})$ on the state $\sigma_t(\phi=0) =(1+\bar{n})^{-1}\sum_{n=0}^{\infty}\left(\frac{\bar{n} }{\bar{n} +1}\right)^{n}|n_{t}\rangle \langle n_{t} |$ has a variance in position  
\begin{equation}
\Delta x_{t}=\langle \hat{x}^{2}\rangle -\langle \hat{x} \rangle^{2}=\frac{\hbar}{2m\omega_{0}}(2\bar{n}+1)e^{-2r_{t}}.\label{eq:variance}
\end{equation}
Using the link to the time-dependent harmonic oscillator through $r_{t}=\ln(\sqrt{\omega_{t}/\omega_{0}})$ yields $\Delta x_{t}=\frac{1}{2k^{2}_{t}\tanh(\varepsilon_{t}/2)}$ with $k_{t}=\sqrt{\frac{m\omega_{t}}{\hbar}}$---see App. \ref{App:Wigner} for details. So  open protocols allow controlling the variance with two parameters, $(\omega_t, \varepsilon_t)$. Figure  \ref{variancemap} shows the variance ratio between final and initial states as function of the inverse temperature and trap frequency relative changes. While the variances at reach with unitary protocols are limited to those on isentropic lines, the map is extended to arbitrary values (including beyond the 3dB limit) thanks to changes in  entropy.

Let us now detail the open control protocols. When  entropy is allowed to change---$\varepsilon_t$ is time-dependent---direct derivation of the state matrix $\sigma_t=\frac{1}{Z_t}e^{-\varepsilon_t (a\dg_t a_t + \frac{1}{2})}$  yields 
 \begin{equation} \label{eq12}
\dot{\sigma}_t = -\frac{i}{\hbar}[H_{t}+H^{(0)}_{\textsc{cd}}, \sigma_t] - \dot{\varepsilon}_t \sigma_t \left( a_t\dg a_t +  \frac{1}{1- e^{\varepsilon_t}}\right),
\end{equation}
with $H^{(0)}_{\textsc{cd}}$ given in Eq. (\ref{H30}). 
The system is initialized in the thermal state (\ref{sigmat}) and evolves according to \eqref{eq12}. As in the unitary case, for the sake of experimental implementation, we consider the state matrix $\varrho_t = U_\Omega \sigma_t U_\Omega\dg$ rotated by the unitary $U_\Omega = \exp\big({ i \frac{\Omega_t}{4\omega_t}\big(a_t + a_t\dg\big)^2}\big)$, with  $H_{\varrho} \equiv U_{\Omega}(H_t+H_{\textsc{cd}}^{(0)})U_{\Omega}\dg + i \hbar \dot{U}_{\Omega} U_{\Omega}\dg$. A first master equation readily follows as  
\begin{equation}\label{ME2}
\dot{\varrho}_t = - \frac{i}{\hbar} [H_{\varrho}, \varrho_t] + \mathcal{D}_{\textsc{cd}}(\varrho_t),  
\end{equation}
where all terms accounted for population changes are in the counter-diabiatic dissipator, $\mathcal{D}_{\textsc{cd}}(\varrho_t) =\sum_n \dot{p}_{n,t} U_\Omega \ket{n_t}\bra{n_t} U_\Omega\dg$. 

Alternatively, part of the counter-diabatic Hamiltonian can be written as a `control' harmonic oscillator, $H_c$, with a `control' frequency chosen of the form of the closed results (\ref{weff}), i.e. $\omega_{c}^2 \equiv \omega_t^2 - \Omega_t^2 - \dot{\Omega}_t$. Explicitly, this yields  
\begin{align} 
H_{\varrho} =&  H_c + \frac{ \hbar}{4 \omega_t} \left( 2\Omega_t^2+ \Omega_t \frac{\dot{\omega}_t}{\omega_t} \right)(a_t+a_t\dg)^2 \nonumber \\
&+ i \hbar \left(\frac{\Omega_t}{2} + \frac{\dot{\omega}_t}{4 \omega_t}\right)(a_t^2 - a_t^{\dagger 2})
\end{align}
with $H_c = \hat{p}^2 / (2m) + \frac{1}{2} m \omega_c^2 \hat{x}^2 = \hbar \omega_t (a_t \dg a_t + \frac{1}{2}) + \frac{\hbar}{4 \omega_t} \omega_c^2 (a_t + a\dg_t)^2.$
Then, the frequency  $\Omega_t \equiv \Omega_0 + \Omega_1$ is taken to be  composed of $\Omega_0 =- \dot{\omega}_t / (2\omega_t)$---to cancel the term in $(a^2_t - a^{\dagger 2}_t)$ if the dynamics were unitary (cf. Eq. \ref{eq1})---and an additional frequency $\Omega_1$ that accounts for changes due to the open dynamics.  
The master equation (\ref{ME2}) thus becomes
\begin{equation}\label{ME3}
\dot{\varrho}_t = - \frac{i}{\hbar} [H_c, \varrho_t] + \mathcal{D}_{c}(\varrho_t). 
\end{equation}
The `control' dissipator can be written in a compact form by defining the annihilation operator 
$ 
b_t \equiv U_\Omega a_t U_\Omega\dg = a_t - i \frac{\Omega_t}{2 \omega_t}(a_t + a_t\dg), 
$
which gives (see App. \ref{app:Dc} for details)
\begin{equation} \label{eq:Dc}
\mathcal{D}_{c}(\varrho_t){=} \frac{\Omega_1}{2}\big[b_t^2 {-} b_t^{\dagger 2}, \varrho_t \big]{-}\dot{\varepsilon}_t \varrho_t \Big( b\dg_t b_t {+} \frac{1}{1{-}e^{\varepsilon_t}}\Big).
\end{equation}
Note that $a_t + a\dg_t = b_t + b_t\dg$, so the position operator $\hat{x}$ is equivalently written in one basis or the other.

This dissipator can be further written in a more `experimentally-friendly' form. For this, notice that the relations $a_t \sigma_t = e^{-\varepsilon_t(a_t a\dg_t+1/2)} a_t Z_t^{-1} =  \sigma_t a_t e^{-\varepsilon_t}$ and $a_t\dg \sigma_t = \sigma_t a_t\dg e^{\varepsilon_t}$ translate to  $b_t \varrho_t = \varrho_t b_t e^{-\varepsilon_t}$ and $b_t\dg \varrho_t = \varrho_t b_t\dg e^{\varepsilon_t}$. By setting 
\begin{equation}\label{Omega1}
\Omega_1 \equiv \frac{\dot{\varepsilon}_t}{(1 - e^{\varepsilon_t})(1 + e^{-\varepsilon_t})}=-\frac{\dot{\varepsilon}_{t}}{2\sinh(\varepsilon_{t})},
\end{equation}  the dissipator \eqref{eq:Dc} can be recast as 
\begin{eqnarray} \label{eq:MExx}
\mathcal{D}_{c}(\varrho_t)&=& - \Gamma_t [(b_t + b\dg_t),[(b_t + b\dg_t),\varrho_t]] \\
 &=&- \gamma_t [\hat{x},[\hat{x},\varrho_t]],  \nonumber
\end{eqnarray}
where  $\Gamma_t \equiv \frac{\dot{\varepsilon}_t}{2 (1 - e^{\varepsilon_t})(1 - e^{-\varepsilon_t})}  = \frac{\hbar}{2 m \omega_t}\gamma_t$. 
The control dissipator  thus becomes the well-known form of localization in the position eigenbasis, often referred to as Joos-Zeh term \cite{joos1985,SchlosshauerBook}, and  is easily implementable in current experimental platforms.  
In turn, the modulation of $\gamma_t$ can be engineered, e.g.,  by post-selection measurement of the position or  via stochastic parametric driving, as proposed in \cite{dupays2020} and also used below in Section \ref{subsec:trap}.

\begin{figure}
\centering
\includegraphics[width=0.85\columnwidth]{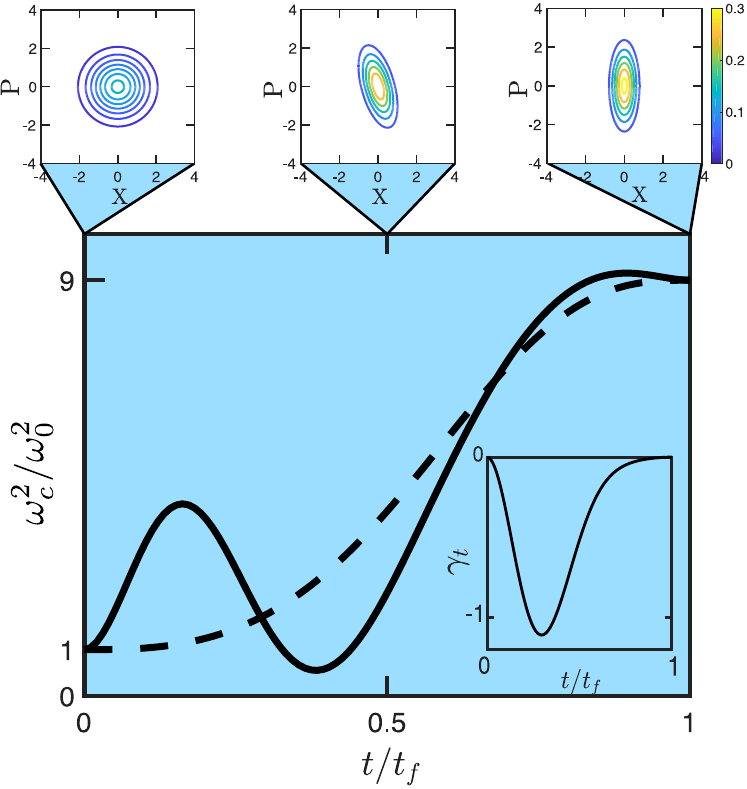}
\caption{Control frequency $\omega^{2}_{c}/\omega^2_0$ and reference Ansatz $\omega_t^2/\omega_0^2$ (dashed line) for the initial conditions $\beta_{0}=\omega_{0}=1$ and final conditions $\beta_{f}=2$ and $\omega_{f}=3$  at $t_f=2$. The Wigner function is plotted at times $t=0$, $t_{f}/2$, and $t_{f}$.  It starts as a symmetric Gaussian and rotates in phase-space during the process to reach a squeezed state along the $\hat{x}$ quadrature, as expected for compression. The inset shows the dissipation rate $\gamma_t$. \label{fig:fq_wig}}
\end{figure}

The designed master equation of interest thus reads 
\begin{equation}
\frac{d \varrho_t}{dt} =\frac{1}{i\hbar}\left[\frac{\hat{p}^{2}}{2m}+\frac{1}{2}m\omega^{2}_{c}\hat{x}^{2},\varrho_t\right]-\gamma_{t}[\hat{x},[\hat{x},\varrho_t]].
\end{equation}
with the control parameters 
\begin{subequations}
\label{eq:control}
\begin{align}
\omega^{2}_{c}&=\omega^{2}_{t}-(\Omega_{0}+\Omega_1)^2-\dot{\Omega}_{0}-\dot{\Omega}_{1} \label{eq:fq_open},\\
\gamma_{t}&=-\frac{m\omega_{t}}{\hbar}\frac{\dot{\varepsilon}_{t}}{4\sinh^{2}(\varepsilon_{t}/2)}.
\end{align}
\end{subequations}
It allows generating the target state $\varrho_f = \sigma_f$ at final time---since then $\dot{\omega}_{f}=\dot{\varepsilon}_{f}=0$.
Implementation easily follows from knowledge of the control parameters (\ref{eq:control}):  given the boundary conditions $\beta_{0},\beta_{f}$ and $\omega_{0},\omega_{f}$, one can fixe the time evolution as e.g. a fifth-order polynomial Ansatz, $p(\tau)=10\tau^{3}-15\tau^{4}+6\tau^{5}$, on $\beta_{t}$ and $\omega_{t}$. Specifically  $\omega_{t}=\omega_{0}+(\omega_{f}-\omega_{0})p(t/t_{f})$ and $\beta_{t}=\beta_{0}+(\beta_{f}-\beta_{0})p(t/t_{f})$. 
With this choice, we illustrate the dynamics in Figure \ref{fig:fq_wig}, which shows the control parameters and the state Wigner function along the open dynamics. As expected for a compression process ($\omega_{f}=3\omega_{0}$), the Wigner representation of the final state evidences a thermal state squeezed in position.

 To summarize this part, we showed how opening the dynamics influences the control frequency---compare Eq. (\ref{eq:fq_open}) with (\ref{weff}). The particular choice of the correction $\Omega_1$ made in Eq. (\ref{Omega1}) leads to a dissipator controled in position, as proposed in \cite{alipour2020,dupays2020}. Should this dissipator not be the most experimentally suited, Eq. (\ref{eq:Dc}) provides the more general result for the control dissipator. Also note that the dynamics is resolved at the level of operators and not restricted to the coordinate representation.  As such, the results here are more general and complements these previous works.

 \section{Squeezing and Thermalization with two-photon Raman interaction \label{sec:2photon}}

A squeezing protocol alternative to controlling the trap frequency is based on two-photon Raman interaction, that was successfully used to squeeze the ground vibrational of trapped ions \cite{Meystre1982,PhysRevA.37.3175,gerry2004}. We detail below the experimental setup used for implementation of equation (\ref{eq:dilatation}), and extend the known technique to allow for (i)  squeezing in arbitrary time thanks to reverse engineering, and (ii) at arbitrary temperature with engineered dephasing. We notably explain how the modulation of the lasers amplitudes allow to modify the variance \eqref{eq:variance} in arbitrary time.

\subsection{Experimental setup \label{sec:exp}}

\begin{figure}
 \begin{center}
\includegraphics[width=0.5\columnwidth]{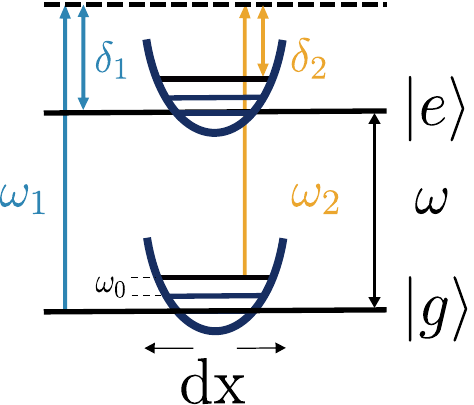}
\caption{Proposed setup for implementation of two-photon Raman interaction with a control dissipator generated by stochastically shaking the trap. \label{fig:scheme1}}
\end{center}
\end{figure}

Consider a trapped ion
 interacting with two mono-chromatic laser beams---see Fig.~\ref{fig:scheme1}. 
 In the experimental situation of interest, the electronic structure of the ion is reduced to a  two-level system  described by the atomic Hamiltonian $H_{a}= \frac{\hbar \omega}{2} \sigma_z$, with $\sigma_z = |e\rangle\langle e| - \ket{g}\bra{g}$.
The motion of the trapped atom can be considered harmonic in all three dimensions, as obtained either from a classical or quantum-mechanical treatment \cite{glauber1992, leibfried2003}, and thus described by $H_m = \hbar \omega_{0} (a\dg a + 1/2)$.  With suitable electromagnetic fields, the electronic levels can be coupled to each other and to the vibrational motional degrees.  Each of the electromagnetic field is treated as a classical plane wave  of the form, in the direction ${\bf x}$ of interest, ${\bf E}_l(\hat{x},t) \cdot {\bf x} = A_l(t)   (e^{i (k_l \hat{x} - \omega_l t- \Phi_l) }+ {\rm c.c.})/2$ with time-dependent amplitude $A_l(t)$, wave vector ${\bf k}_l = k_l {\bf x}$, and detuning $\delta_l$ from the atomic transition, $\omega_l - \delta_l = \omega$. The interaction Hamiltonian resulting from the applied two laser fields can be described as \cite{leibfried2003}
\begin{equation}
H_{\rm int}(t) = {\sum_{l=\{1,2\}} } \frac{\hbar}{2}\Omega_l \sigma_x\left( e^{i (k_l \hat{x} - \omega_l t - \Phi_l)} {+} {\rm h.c.}\right),
\end{equation}
with $\sigma_x = \ket{g}\bra{e} + \ket{e}\bra{g}$. 
The Rabi frequency describing dipole coupling  to a single charge $q$  is given by $\hbar \Omega_l /2 = q  \bra{g} \hat{x} \ket{e} A_l(t)  $.

We aim at preparing a squeezed thermal state on the vibrational levels of the system with total Hamiltonian
 \begin{equation}
\label{eq:eqtot}
H_{\rm tot}(t)=H_{\rm a}+H_{\rm m}+H_{\rm int}(t),
\end{equation}
starting from an initial vibrational state that is thermal.

\subsection{Closed dynamics}
We first consider the unitary dynamics and denote  $\ket{\psi_t}$  the solution of the Schr\"odinger equation.  It is useful to change the energy scale \cite{brion2007} and look at the evolution of $\ket{\Psi_t} = U_{\rm r,t} \ket{\psi_t}$ rotated by a unitary  transformation $U_{{\rm r},t}\equiv e^{\frac{i}{\hbar} H_{\rm r} t}$. The rescaling Hamiltonian $H_{\rm r}=H_{\rm a} +H_{\rm m} +  \frac{\hbar \Delta}{2}\sigma_z$ effectively shifts the electronic energy  of the gap $\hbar\omega$ into an energy defined by the average laser detuning, $\hbar \Delta=\hbar(\delta_{1}+\delta_{2})/2$, and yields to an interaction picture. The rotated state evolves as 
$i \hbar \ket{\dot{\Psi}_t} = H \ket{\Psi_t}$, with $H\equiv U_{\rm r,t} H_{\rm tot} U_{\rm r,t}\dg + i \hbar \dot{U}_{\rm r,t} U_{\rm r,t}\dg$ explicitly reading
\begin{eqnarray}
\label{eq:rotatedhamiltonian}
H&=&- \frac{ \hbar \Delta}{2} \sigma_z +  U_{\rm r,t} H_{\rm int}(t) U_{\rm r,t}\dg \nonumber  \\
&=&{-} \frac{ \hbar \Delta}{2} \sigma_z +\sum_{l=1,2}\frac{\hbar}{2}\Omega_{l} \\
&&\:\times\left(e^{i(\omega_{l}{-}\omega-\Delta)t}e^{i\Phi_{l}}e^{-i\eta_{l}(a_t+a_t\dg)}
|g\rangle\langle e| + {\rm h.c.} \right) .\nonumber
\end{eqnarray}
The position being quantized, we used $\hat{x} = x_0 (a+ a\dg)$ to write $e^{i k_l \hat{x}}=e^{i\eta_{l}(a+a^{\dagger})}$ in the expression of the electromagnetic field \cite{gerry2004}. The interaction picture leads to using the time-dependent operators $a_t \equiv U_{{\rm r},t}a U_{{\rm r},t} = a e^{- i \omega_{0} t}$ and $\hat{x}_t \equiv U_{{\rm r},t}\hat{x}U_{{\rm r},t} = x_0(a_t + a\dg_t)$.  The Lamb-Dicke parameter $\eta_{l}= k_l x_0$ is defined from the extension of the ground-state wave function of the reference oscillator, $x_0 = \sqrt{\hbar/(2m \omega_{0})}$. 
The evolving wave function is a superposition of the electronic ground and excited states dressed with the vibrational levels $\ket{n}$, and we look for a solution in the form 
$|\Psi_{t}\rangle=\sum_{n}\big( g_{n}(t)|g,n\rangle+ e_{n}(t)|e,n\rangle\big)$.  The electronic and vibrational degrees of freedom can be decoupled through an adiabatic elimination \cite{brion2007}, which assumes constant excited-state population. We follow \cite{muga2010a, lizuain2008} and set $\omega_1 - \omega_2 = 2\omega_{0}$. Keeping only the resonant, second blue sideband, which effectively is a vibrational form of the RWA, and neglecting the Lamb term shifting, the evolution of the atom state density is dictated by 
the effective squeezing Hamiltonian as computed in App. \ref{app:unitary}
\begin{equation}
\label{eq:heff}
H_{\rm eff}(t)=\hbar (\eta_{2}{-}\eta_{1})^{2}\frac{\Omega_{1}\Omega_{2}}{4\Delta}\big(e^{i(\Phi_{1}{-}\Phi_{2})} a^{2}{+}{\rm h.c}\big)|g\rangle\langle g|.
\end{equation}

One can rewrite the evolution of $|\Psi_{t}\rangle$ in the Liouville-von Neumann form, so that $\rho_{t}=|\Psi_{t}\rangle \langle \Psi_{t}|$ evolves as 
\begin{align} \label{eq:mastereqrot}
\frac{d\rho_t}{dt}
&=-i [\alpha_t a^{2}+ \alpha^{*}_t a^{\dagger 2},\rho_t], 
\end{align} 
with 
\begin{equation}\label{eq:alpha}
\alpha_t =(\eta_{2}-\eta_{1})^{2}\frac{\Omega_{1}(t)\Omega_{2}(t)}{4\Delta}e^{i\Phi_t}
\end{equation}
and $\Phi_t = (\Phi_{1}-\Phi_{2})$.  Choosing the dephasing between the lasers  to be $\Phi_{1}-\Phi_{2}=\frac{\pi}{2}$ recovers the 2-photon Raman Hamiltonian. 

 \begin{figure}
 \centering
 \includegraphics[width=1\columnwidth]{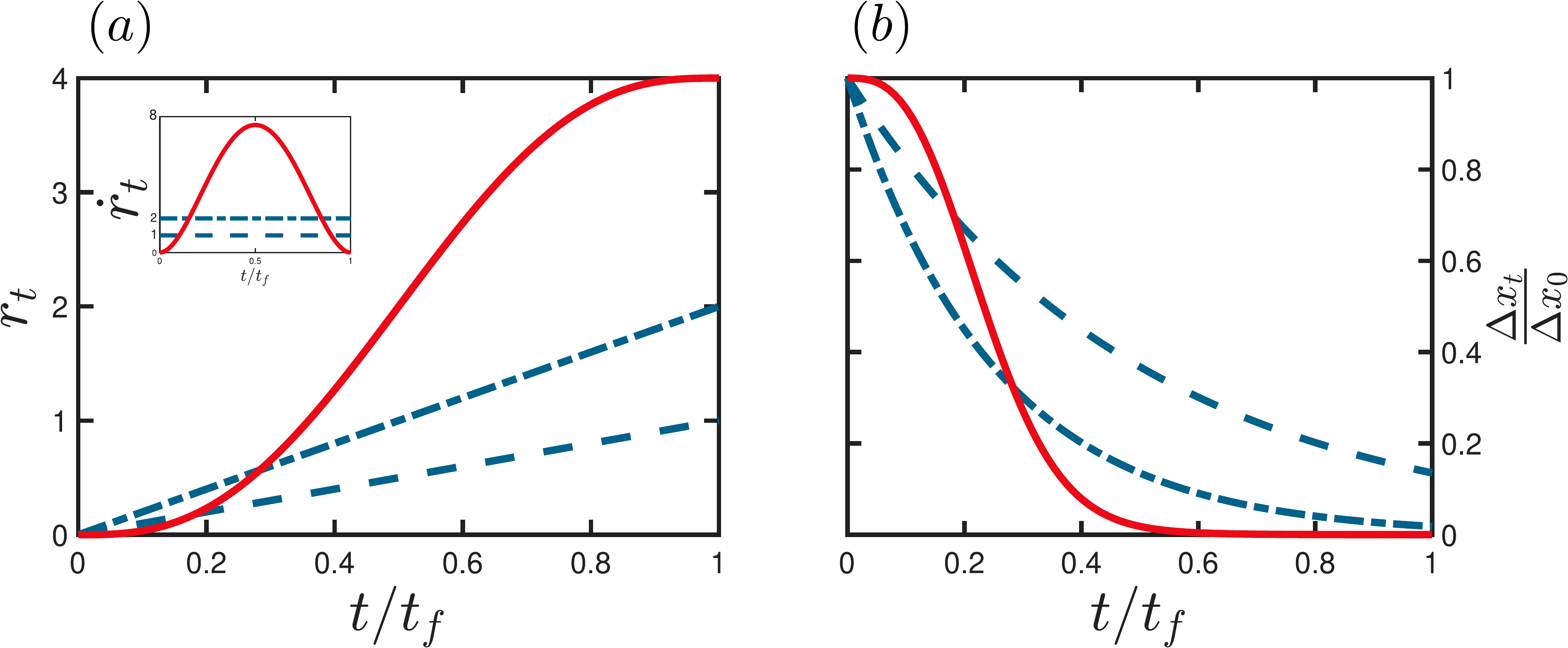}
\captionof{figure}{ Evolution of (a) {\bf Squeezing amplitude} $r_t$ as function of the process time with linear variation  (blue curves) or  through a  controlled dynamics (red curve);  and corresponding (b) {\bf normalized  variance in position},  Eq. (\ref{eq:variance}). 
Increasing the squeezing parameter linearly in time ($\dot{r}_t$ constant, see inset) yields to a variance that decreases exponentially in time. STA techniques, through reverse-engineering of the dynamics, allow to reach a target squeezing amplitude in a controlled time. For example, taking  $r_\tau=r_{0}+(r_f-r_{0})(10\tau^{3}-15\tau^{4}+6\tau^{5})$ with $\tau=t/t_{f}$ leads to a desired squeezing in arbitrary time ($r_{f}=4$ and $t_f=1$ here). 
\label{explanation}}
\end{figure}

Importantly, this recovers the dynamics of the squeezed state given by (\ref{eq:dilatation}) in the rotated frame, thus generating the general squeezing Hamiltonian \eqref{H3} in the original frame. In this setup,  the squeezing parameter is directly related to the  experimental parameters as 
\begin{equation}
\frac{\dot{r}_{t}}{2} = |\alpha_t|. 
\label{eq:simple_but_nice}
\end{equation}
This simple relation is experimentally very relevant. It implies that using a constant amplitude for the lasers leads to a  squeezing parameter linear in time. So to achieve a fixed squeezing parameter, one must wait a given time. By contrast, the same squeezing parameter can be achieved in an arbitrary time through reverse engineering. For example, consider a fifth-order polynomial interpolating between the initial and final squeezing parameter, $r_t=r_{0}+(r_f-r_{0})p(t/t_f)$. Its derivative gives, through Eq. \eqref{eq:simple_but_nice}, the laser amplitudes needed to reach the target squeezing in arbitrary time. This is further illustrated in Fig. \ref{explanation}, that shows the benefit of reserve engineering the laser amplitudes over using a constant amplitude in order to reach a target squeezing parameter in a desired time. Furthermore, as mentioned in Sec. \ref{sec:thermalstate}, the time of the process determines the phase. It is then possible to choose the time of the process accordingly to the desired target phase,  $t_{f}=-\frac{\phi_{f}}{2\omega_{0}}$.
 We now extend the dynamics to open processes and solve the state evolution. In addition to provide control over the temperature, the derived solution allows generating an arbitrary squeezing phase.

\subsection{Open dynamics: Engineering the master equation in a stochastically-shaken trapped ion \label{subsec:trap}} 

In order to generate a controlled dissipation and extend the map of final states at reach, we process as in the HO, and use stochastic processes. Specifically, we add to the total Hamiltonian a stochastic component, and consider
\begin{equation}\label{eq:hst}
h_{\rm st}=H_{\rm tot}(t)+\hbar\sqrt{2\gamma_{t}}\xi_{t}\hat{x}\otimes|g \rangle \langle g |. 
\end{equation} 
This system is now characterized by the Wiener process $W_{t}=W_{0}+\int_{0}^{t}\xi_{t'}dt'$ defined in terms of the  normally distributed random variable $\xi_{t}$ taken as white noise---with zero mean and vanishing correlation, $\langle \xi_t \xi_{t'} \rangle = \delta(t-t')$. 
The stochastic term allows to create the control dissipator and has the  advantage of being readily implementable via  continuous quantum measurement or in a stochastically-shaken trap \cite{jacobs2006}. Note that the stochastic term in \eqref{eq:hst} is taken as acting only on the electronic ground state. 
While this yields a rigorous analytical derivation, it can be experimentally challenging, so an alternative scheme is presented in App. \ref{sec:laser} that, instead of shaking the trap, relies on two additional laser beams, one having a stochastic amplitude.

Let $\ket{\psi_t}$ denote the solution of the Schr\"odinger equation, $\ket{\psi_{t+dt}} = e^{-\frac{i}{\hbar}h_{\rm st} dt}\ket{\psi_t}$. 
Following the unitary results, we look at the  evolution of the state vector $\ket{\Psi_t} = U_{\rm r,t} \ket{\psi_t}$. The influence of the additional stochastic term is detailed in App.  \ref{app:trap}.
The density matrix of interest is the ensemble one, obtained from averaging over the realizations of the noise and    denoted $\rho_t= \la  \ket{\Psi_t}\bra{\Psi_t}\ra$. 
We find its evolution dictated by the master equation 
\begin{align} \label{eq:mastereqrot}
\frac{d\rho_t}{dt}&=-\frac{i}{\hbar}[H_{\rm eff}(t),\rho_t]-\gamma_{t}[\hat{x}_t,[\hat{x}_t,\rho_t]]  \\
&=-i [\alpha_t a^{2}+ \alpha_t^* a^{\dagger 2},\rho_t]+2\kappa_t\left(\mathcal{D}(a)+\mathcal{D}(a^{\dagger})\right) \nonumber, 
\end{align} 
 the second line following from the RWA. It corresponds to a master equation of Lindblad form, where  
 the dissipators, defined from  $\mathcal{D}(a)=a\rho_t a^{\dagger}-\frac{1}{2}\{a^{\dagger}a,\rho_t\}$, are modulated with an amplitude $\kappa_t = \gamma_t x_0^2 $. The parameter $\alpha_t  $ is the same as in the unitary case---Eq. (\ref{eq:alpha}). We solve the dynamics and show how this setup can be used to generate a target squeezed thermal state.

\subsection{Solving the dynamics}
The system is initialized in a state with density matrix $\ket{g}\bra{g} \otimes \sigma_0$, where $\sigma_t \equiv e^{- \beta_t H_m}/ Z_t$ denotes the thermal state on the vibrational manifold at initial inverse temperature $\beta_0$. 
We next solve the dynamics to find the dynamical  control parameters $\{\alpha_t, \kappa_t\}$ for which the  squeezed thermal state 
\begin{align}
\label{eq:solution}
\rho_t&=|g\rangle\langle g| \otimes S_{r, \phi} \sigma_0 S^{\dagger}_{r, \phi}\nonumber \\
&=|g\rangle\langle g| \otimes \frac{1}{Z_{t}} e^{\lambda_t \left( S_{r, \phi}a^{\dagger}aS^{\dagger}_{r, \phi} + \frac{1}{2}\right)}
\end{align}
is solution of  the dynamics (\ref{eq:mastereqrot}). The time-dependent parameter $\lambda_t \equiv - \beta_t \hbar \omega_{0}$  allows varying the temperature. It is useful to work with the factorized form, that we derive in normal ordering following McCoy \cite{mccoy1932a} as (see App.  \ref{app:fact})
\begin{align}\label{eq:facto}
&e^{\lambda S_{r, \phi}a^{\dagger}aS\dg_{r, \phi}} \nonumber\\ 
&= e^{\lambda(\cosh(2r) a\dg a {+} \cosh r \sinh r (e^{i \phi} a^{\dagger 2} {+} e^{{-} i \phi}  a^{2}) {+} \sinh^2r \mathbbm{1})} \nonumber \\
&= K_t e^{J^{*}_t a^{\dagger 2}}e^{-B_t a^{\dagger}a}e^{J_t a^{2}},
\end{align}
where the parameters are defined as $J_t = j(r_t, \lambda_t)e^{ i \phi_t}$ with the real functions $j(r, \lambda) \equiv \frac{1}{2}\left(\frac{\sinh(2r)(e^{2\lambda}-1)}{2(\cosh^{2}r-\sinh^{2}re^{2\lambda})}\right)$,  $B_t \equiv -\ln\left(\left|1+\frac{(e^{\lambda_t}-1)(\cosh^{2}(r_{t})+\sinh^{2}(r_{t})e^{\lambda_t})}{\cosh^{2}(r_{t})-\sinh^{2}(r_{t})e^{2\lambda_t}}\right|\right)$, and the normalizing constant $K_t$---given explicitly in Eq. (\ref{eq:K}).
The master equation (\ref{eq:mastereqrot}) gives 
\begin{align} \label{eq:rhoirho}
\frac{d \rho_t}{dt} \rho_t^{-1}&= - i \alpha_t^* a^{\dagger 2} - i \alpha_t a^2 - 2 \kappa_t - 2 \kappa_t a\dg a \nonumber \\
&+ \rho_t \left(i \alpha_t^*a^{\dagger 2}  + i \alpha_t a^2 - 2 \kappa_t a\dg a\right) \rho_t^{-1} \nonumber \\
 &+2\kappa_t a \rho_t a\dg \rho_t^{-1} + 2\kappa_t a^{\dagger} \rho_t a \rho_t^{-1}. 
\end{align}

Using the adjoint representation, detailed in App. \ref{app:trap}, yields to the simple system linking the vector of the control parameters $v_{c}=(\kappa \quad \alpha_{R} \quad \alpha_{I})^{T} $ to the vector of the squeezing parameters $v_{sq}=(\dot{J}_{R} \quad \dot{J}_{I} \quad \dot{B})^{T}$ with $^{T}$ the transposition.  Specifically, we obtain 
 \begin{align}
\label{eq:dyneq1}
v_{c}=M_{t}^{-1}v_{sq}
\end{align}
 with the transfer matrix
  \vspace*{-0.2cm}
 \onecolumn
\setlength{\arraycolsep}{1pt} 
\begin{align}\label{eq:M1}
M_{t}=
\begin{pmatrix}
4(e^{-B}-1)J_{R}&-8J_{I}J_{R}&4(J^{2}_{R}-J^{2}_{I})+(e^{-2B}-1)\\
4(e^{-B}-1)J_{I}&4(J^{2}_{R}-J^{2}_{I})+(1-e^{-2B})&8J_{R}J_{I}\\
-4\left(\cosh B-1+2e^{B}(J^{2}_{R}+J^{2}_{I})\right)&8J_{I}&-8J_{R}\\
\end{pmatrix}.
\end{align}
\twocolumn
\noindent 
This is the main result of this section. Eqs. (\ref{eq:dyneq1} - \ref{eq:M1}) give the control parameters $\alpha_t = |\alpha_t| e^{i \Phi_t} = \alpha_R + i \alpha_I$ and  $\kappa_t= \gamma_t x_0^2$ to engineer the squeezed state characterized by 
$J_t = J_R + i J_I$ and $B_t$ at the desired temperature through $\lambda_t \equiv -\beta_t\hbar \omega_{0} $.

We show numerical applications of the controlled parameters to be implemented to drive an initial (possibly squeezed) thermal state characterized by $\{r_0, \phi_0,\lambda_0\}$ into a target final state $\{r_f, \phi_f,\lambda_f\}$. The state parameters $J_{R},J_{I}$ and $B_t$ are assumed to follow a smooth evolution taken as a fifth-order polynomial, with additional  boundary conditions taken as null  first and second derivatives at initial and final times. 
 The relative detuning between the lasers is fixed to $\omega_{2}-\omega_{1}=2\omega_{0}$, as required to generate the squeezing Hamiltonian (\ref{eq:heff}). The control parameters are  obtained by solving Eqs. (\ref{eq:dyneq1}, \ref{eq:M1}). The dynamics can thus be implemented through the controlled dephasing strength $\kappa_t = \gamma_t x_0^2$, the controlled laser amplitudes, and their Rabi frequencies.  The latter are 
  directly related to the control parameters  $\alpha = |\alpha| e^{i(\Phi_{1}-\Phi_{2})}$ that gives the relative laser phases $\Phi_1 - \Phi_2 = \arctan\left(\alpha_{I}/\alpha_{R}\right)$ and Rabi frequencies through $|\alpha| = (\eta_{2}-\eta_{1})^{2}\frac{\Omega_{1}\Omega_{2}}{4\Delta}$.
  
 \begin{figure}
\includegraphics[width=\columnwidth]{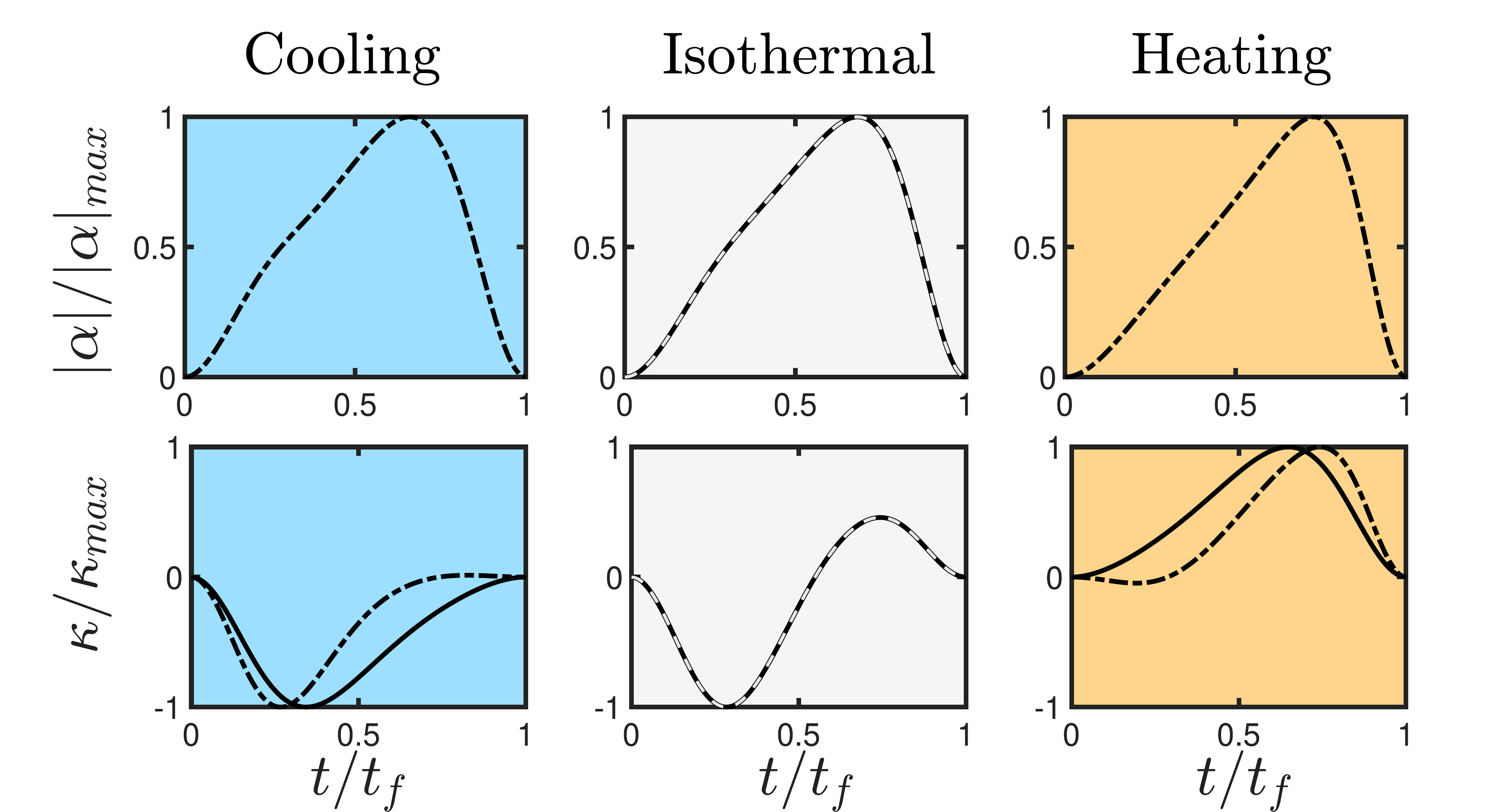}
\captionof{figure}{{\bf Experimental control parameters}: (top) laser relative amplitude and (bottom) dephasing strength for (a) cooling ($\lambda_f=-2$), (b) isothermal ($\lambda_f=\lambda_i$), and (c) heating ($\lambda_f=-0.5$) processes. The initial state is isotropic  $r_{i}=\phi_{i}=0$ at $\lambda_i=-1$. The final state ($t_f=1$) is a thermal state with 
no squeezing  $r_{f}=\phi_{f}=0$ (plain lines); squeezing at $r_{f}=1, \phi_f =0$ (dash-dotted lines),  or  squeezing at $r_{f}=1$ and angle $\phi_{f}=\frac{\pi}{4}$ (grey dashed lines). The control parameters are normalized---see Figs. \ref{fig2}-\ref{fig3} for the influence of temperature and squeezing on their maxima.\label{fig:control2laser}}
\end{figure}

\begin{figure}
\includegraphics[width = 1.\columnwidth]{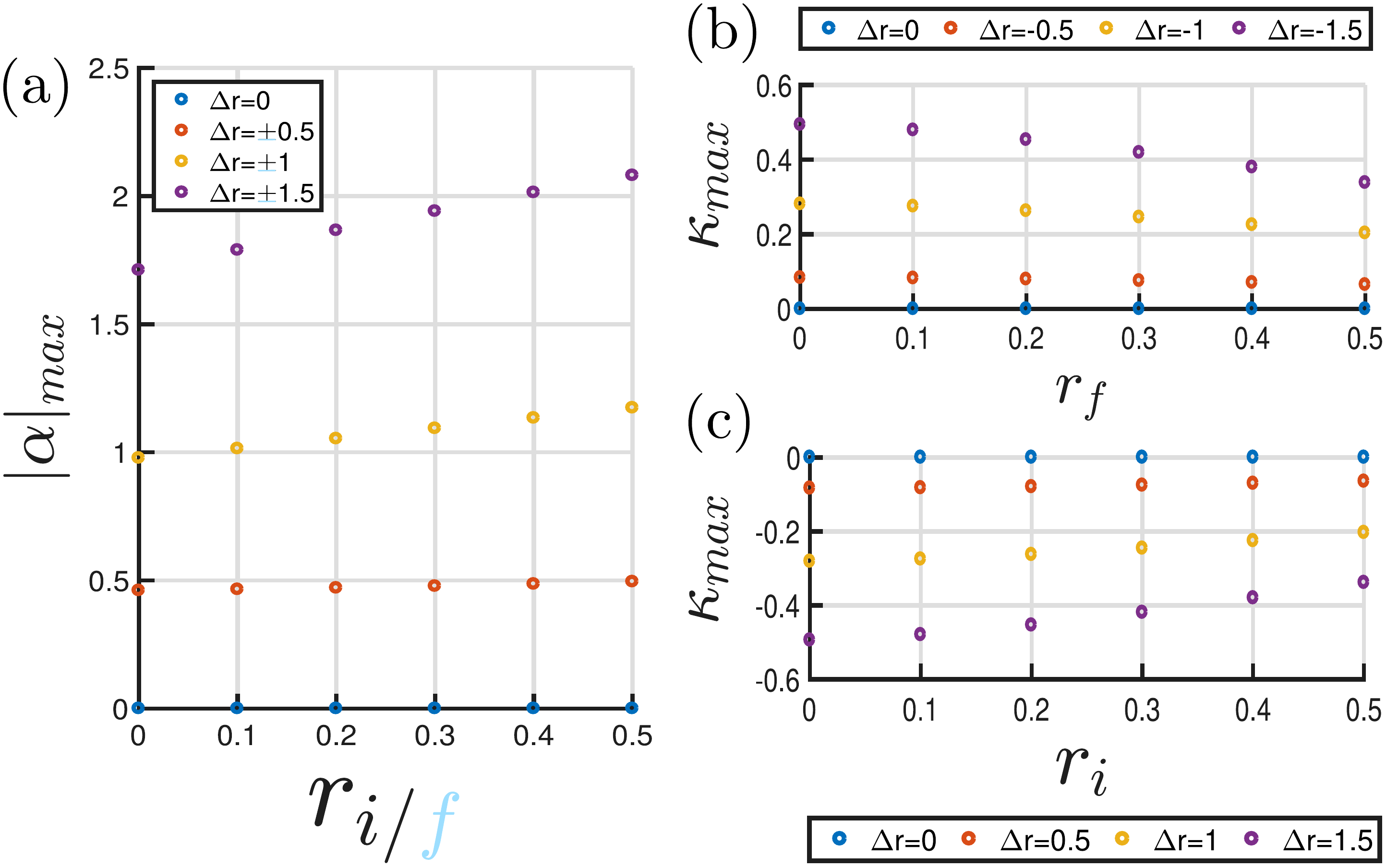}
\captionof{figure}{{\bf Influence of squeezing amplitude on the control maxima}: Maximum (a-b) dephasing strength and (c) laser amplitude as function of the initial $r_i$ or final $r_f$ squeezing amplitude for different variation of squeezing $\Delta r = r_f - r_i$. Plots are for $t_f=1$. \label{fig3}}
\end{figure}

\begin{figure}
\centering
\includegraphics[width = 1\columnwidth]{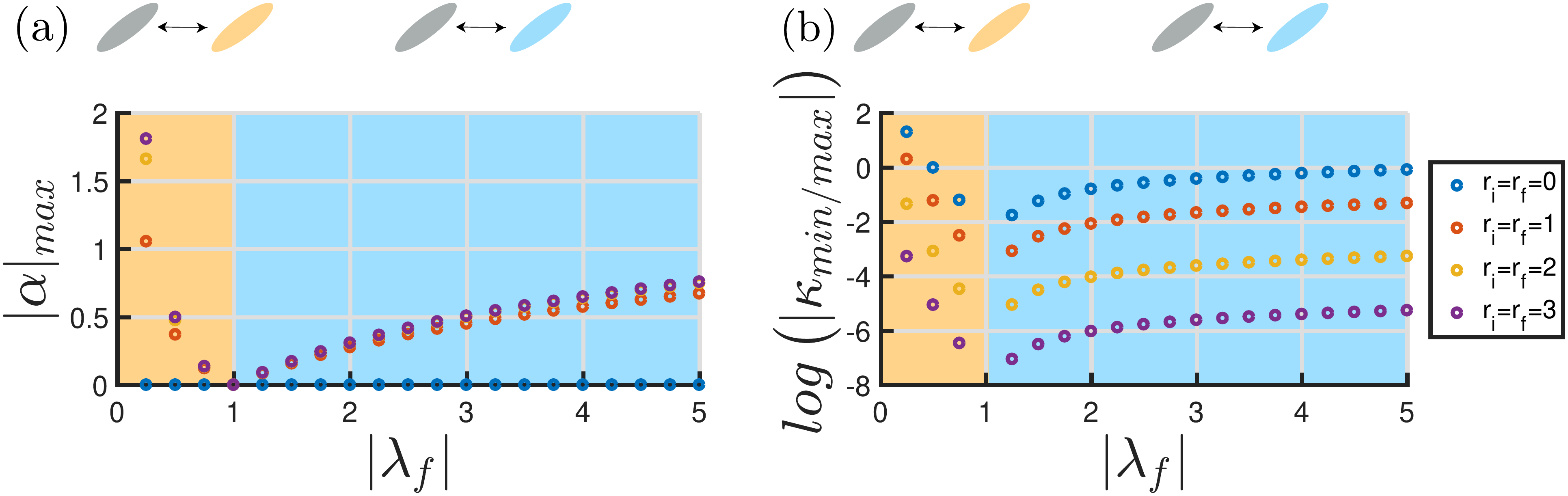}
\captionof{figure}{{\bf Influence of temperature on the control maxima}:  Maximum (a) laser amplitude  and (b) dephasing strength as function of changes in the temperature $|\lambda_f | = \beta_f \hbar \omega_{0}$ for heating (orange background) and cooling (blue background) processes ($t_f=1$). Results are shown for states with constant squeezing amplitude, $r_i=r_f$, starting with $|\lambda_i|=1$ and $\phi_{i}=\phi_{f}=0$.  \label{fig2}}
\end{figure}

 Figure \ref{fig:control2laser} shows the control parameters for squeezing with different temperature conditions, namely cooling, isothermal,  and heating. The normalized laser amplitude appears to be quite similar for all squeezing processes, which can be expected as it mainly controls the squeezing amplitude. In turn, its maximum is influenced by the variation of squeezing, as shown in Fig. \ref{fig3}.  When the state retains an isentropic density ($\Delta r=0$), no squeezing term is needed, as intuitively expected.  Figures \ref{fig3} and  \ref{fig2} show the influence of changing the  squeezing amplitude and  temperature, respectively.  We verify that the dephasing strength is `symmetric in squeezing' in the sense that squeezing by a positive or negative variation $|\Delta r=r_f - r_i|$ only changes the sign of the dephasing, not its strength.

As mentioned above, implementation of the stochastic Hamiltonian (\ref{eq:hst}) assumes a spin-dependent term on the position of the trap, that could be developed following the techniques proposed in e.g. \cite{haljan2005}. This allowed for a rigorous derivation of the effective Hamiltonian through the adiabatic elimination. Shaking the full trap (ground and excited electronic states) would require further approximations of the excited state populations, although the adiabatic elimination might still hold at large detunings. Further work could be done using the recently developed adiabatic elimination for open bipartite systems \cite{saideh2020projection,Finkelstein_Shapiro_2020,Reiter2011}. 
An experimental alternative 
 is to install a feedback loop that enforces the qubit to remain in its ground state \cite{Genoni_2015}. 
Should the proposed model still be  experimentally limiting, we provide in App.~\ref{sec:laser} an alternative scheme where the dissipator is engineered  with two additional laser field instead of shaking the trap.

Finally, note that we have here focused on the trapped-ion setup for the sake of proposing a scheme that can be directly implemented experimentally. However, the dissipator need not be of the form $[\hat{x},[\hat{x},\rho]]$. For example, using a  dissipator of the type $\mathcal{D}_{\rm sq}=\gamma\left(\bar{n}+1\right)\mathcal{L}(a)\rho+\gamma \bar{n}\mathcal{L}(a\dg)\rho$ in Eq. (\ref{eq:mastereqrot}), with $\mathcal{L}(a)=a \rho a\dg-\frac{1}{2}\left(a\dg a \rho+\rho a\dg a\right)$, could also be used to generate a squeezed thermal state in a photonic platform, where the squeezing Hamiltonian could be obtained with, e.g., parametric downconversion.

\bigskip

\section{Conclusion \label{sec:conclu}}
Starting from the general  evolution for a squeezed thermal, we first  clarified how squeezing without phase control can be achieved in arbitrary time by modulating the trap-frequency of a harmonic oscillator and as such, relates to known STA techniques. In turn, control of the phase can be implemented with a two-photon Raman Hamiltonian.  Importantly, the two  approaches presented here include dissipative dynamics in order to control the state entropy, that is engineered using stochastic fields. 
We provided a detailed analysis in a trapped-ion setup, giving the control laser amplitude, relative phase, and dephasing strength suited  to generate a target squeezed state in arbitrary time. 
The general formalism  could also capture, e.g., photonic thermal states squeezed by parametric downconversion in a lossy cavity  \cite{seifoory2016} and is thus adaptable to other experimental platforms. 
 Among possible applications, the generated squeezed states can be used for trapped-ion transport \cite{sutherland2021}, which is relevant to quantum computing architectures.

\medskip
\textbf{Acknowledgements} \par 
It is a pleasure to thank Adolfo del Campo, Kihwan Kim, and Mauro Paternostro for insightful discussions and comments on the manuscript. 


\appendix
\onecolumn

\setcounter{equation}{0}
\renewcommand{\theequation}{S\arabic{equation}}

\section{Finding an expression for $S_{r,\phi} \dot{S}\dg_{r, \phi}$ \label{app:SSdot}}
To find an explicit expression of  $S_{r,\phi} \dot{S}\dg_{r, \phi}$  for any squeezing angle $\phi_t$, it is useful to define the operator $\tilde{a}_t = e^{- i \frac{\phi_t}{2}}a_0$. It fulfills the bosonic commutator relation $[\tilde{a}_t, \tilde{a}\dg_t]$, and gives $S_{r,\phi} = e^{\frac{r_t}{2}(\tilde{a}_t^2 - \tilde{a}_t^{\dagger 2})}$. We can expand the exponential in Taylor series and look for $S_{r,\phi} \dot{S}\dg_{r, \phi}$ in terms of the $k$th derivatives of $\tilde{a}_t^2 - \tilde{a}_t^{\dagger 2}$.  From the time derivative $\dot{\tilde{a}}_t = - i \frac{\dot{\phi}_t}{2} \tilde{a}_t $, it follows that  
$\frac{d(\tilde{a}_t^2 - \tilde{a}_t^{\dagger 2})}{dt} = - i \dot{\phi}_t (\tilde{a}_t^2 +\tilde{a}_t^{\dagger 2})$. We use the fact that $ \tilde{a}_t\dg \tilde{a}_t + \tilde{a}_t \tilde{a}\dg_t= -\frac{1}{4} [\tilde{a}_t^2 +\tilde{a}_t^{\dagger 2},\tilde{a}_t^2 - \tilde{a}_t^{\dagger 2}]$ and $\tilde{a}_t^2 + \tilde{a}_t^{\dagger 2} = \frac{1}{2} [\tilde{a}_t^2 - \tilde{a}_t^{\dagger 2},\tilde{a}\dg_t \tilde{a}_t]$ to  obtain the expression for $k=1$ in the form of commutators, explicitly, 
\begin{equation}
\frac{d(\tilde{a}_t^2 - \tilde{a}_t^{\dagger 2})}{dt} = i \frac{\dot{\phi}_t}{2}[\tilde{a}\dg_t \tilde{a}_t, \tilde{a}_t^2 - \tilde{a}_t^{\dagger 2} ].
\end{equation}
This form allows to generalize the results and obtain 
\begin{equation}
\frac{d(\tilde{a}_t^2 - \tilde{a}_t^{\dagger 2})^k}{dt} =i \left[  \frac{\dot{\phi}_t}{2}\tilde{a}\dg_t \tilde{a}_t  , (\tilde{a}_t^2 - \tilde{a}_t^{\dagger 2})^k\right].
\end{equation}
Eventually, we get  
\begin{eqnarray}\label{SA1}
S_{r,\phi} \dot{S}\dg_{r,\phi} &=& S_{r,\phi} \sum_{k=0}^\infty \frac{1}{k!} \left(\frac{r_t}{2}\right)^k \frac{d}{dt}(\tilde{a}_t^2 - \tilde{a}_t^{\dagger 2})^k \nonumber \\
&=& - \frac{\dot{r}_t}{2}(\tilde{a}_t^2 - \tilde{a}_t^{\dagger 2}) +  i S_{r,\phi} \left[  \frac{\dot{\phi}_t}{2}\tilde{a}\dg_t \tilde{a}_t  , S_{r,\phi}\dg\right]\\
&=& {-} \frac{\dot{r}_t}{2}(e^{- i \phi_t}a_0^2 {-} e^{ i \phi_t}a_0^{\dagger 2}) +  i  \frac{\dot{\phi}_t}{2}(A_t\dg A_t {-} a\dg_0 a_0) , 
\end{eqnarray}
which corresponds to Eq. \eqref{H3} given in the main text.

\section{Wigner function for squeezed thermal state \label{App:Wigner}}
We use the coordinate representation of the state evolving in a  time-dependent harmonic oscillator under dephasing in position  \cite{dupays2020}
\begin{equation}
\langle y | \rho_{t} | x \rangle=\sqrt{\frac{k^{2}_{t}\tanh(\varepsilon_t/2)}{\pi\hbar}}\exp\Big[-\frac{k^{2}_{t}}{2}(y^{2}+x^{2}) \coth \varepsilon_{t}
-i\frac{m}{2\hbar}\Big(\frac{\dot{k_{t}}}{k_{t}}+\frac{\dot{\varepsilon}_{t}}{\sinh(2\varepsilon_{t})}\Big)(y^{2}-x^{2})+\frac{k^{2}_{t}}{\sinh\varepsilon_{t}}yx\Big]\, 
\end{equation}
with $k_{t}=\sqrt{\frac{m\omega_{t}}{\hbar}}$ and $\varepsilon_{t}=\beta_{t}\hbar\omega_{t}$. This readily gives  the Wigner function as
\begin{eqnarray}
W(x,p)&=&\frac{1}{\pi \hbar}\tanh(\varepsilon_{t}/2)
\exp\Big[-\tanh(\varepsilon_{t}/2)\left(k^{2}_{t}+\Big(\frac{m}{k_{t}\hbar}\Big)^{2}\Big(\frac{\dot{k}_{t}}{k_{t}}+\frac{\dot{\varepsilon}_{t}}{\sinh(2\varepsilon_{t})}\Big)^{2}\right)x^{2}\Big]
\nonumber\\
&\times&\exp\Big[-\frac{2m}{\hbar^{2}k^{2}_{t}}\tanh(\varepsilon_{t}/2)(\frac{\dot{k}_{t}}{k_{t}}+\frac{\dot{\varepsilon}_{t}}{\sinh(2\varepsilon_{t})})xp\Big] \exp\Big[-\frac{1}{\hbar^{2}k^{2}_{t}}\tanh(\varepsilon_{t}/2)p^{2}\Big]. 
\end{eqnarray}
Integrating over momenta, one can find the marginal distribution for the position
\begin{equation}
\int dp W(x,p)=\frac{1}{\sqrt{2 \pi \Delta x}}\exp\left(-\frac{1}{2}\frac{x^{2}}{\Delta x}\right) 
\end{equation}
 and identify the variance in position $\Delta x$ as
\begin{equation}
\Delta x_t=\langle x^{2}\rangle -\langle x \rangle^{2}=\frac{1}{2k^{2}_{t}\tanh(\varepsilon_{t}/2)}.
\end{equation}
This result recovers the known variance for a squeezed thermal state \cite{kim1989a}.

\section{Control dissipator $\mathcal{D}_c$ (\ref{eq:Dc})\label{app:Dc}}
The `control' dissipator is defined using Eqs. (\ref{ME2}- \ref{ME3}) as \
\begin{eqnarray}
 \mathcal{D}_{c}(\rho_t) &=& -\frac{i}{\hbar} \left[\frac{\hbar}{4 \omega} \Big(2 \Omega_t^2 + \Omega_t \frac{\dot{\omega}_t}{\omega_t}\Big) (a_t + a\dg_t)^2 + i \hbar \frac{\Omega_1}{2}(a_t^2 - a_t^{\dagger 2}),\rho_t\right] - \dot{\varepsilon}_t \rho_t \left(U_\Omega a_t\dg a_t U_\Omega\dg + \frac{1}{1-e^{\varepsilon_t}}\right) \nonumber . \\
 &=& \frac{\Omega_1}{2}\left[a_t^2 - a_t^{\dagger 2} - i \frac{\Omega_t}{\omega_t}(a_t + a\dg_t)^2 ,\rho_t \right]-\dot{\varepsilon}_t \rho_t \left(U_\Omega a_t\dg a_t U_\Omega\dg + \frac{1}{1-e^{\varepsilon_t}}\right) \label{eq:Dc1}
\end{eqnarray}
In order to find a compact form, it is useful to define the operator 
\begin{equation}
b_t \equiv U_\Omega a_t U_\Omega\dg = a_t - i \frac{\Omega_t}{2 \omega_t}(a_t + a_t\dg).
\end{equation} 
First note that since $a_t + a\dg_t = b_t + b_t\dg$, the position operator is equally represented in both operator basis, namely $\hat{x} = \sqrt{\frac{\hbar}{2 m \omega_t}} (a_t + a\dg_t) = \sqrt{\frac{\hbar}{2 m \omega_t}} (b_t + b\dg_t)$.
Then, noticing that 
$
b_t^2 - b_t^{\dagger 2} = a_t^2 - a_t^{\dagger 2} - i \frac{\Omega_t}{\omega_t}(a_t + a_t\dg)^2, 
$
we can recast  the control dissipator \eqref{eq:Dc1} into the compact form given in Eq. (\ref{eq:Dc}) of the main text.

\section{Factorization of the  squeezed thermal state \label{app:fact}}
It is useful to write the squeezed thermal state in a product form in order to solve its dynamics. We show how to obtain 
\begin{equation} \label{factorization}
e^{\lambda S_{r,\phi} a\dg a S_{r, \phi}\dg} = e^{\lambda A\dg A} = K e^{\frac{X^{*}}{2}a{^{\dagger 2}}} e^{Y a\dg a} e^{\frac{X}{2} a^2}.
\end{equation}
The full analytical demonstration we propose here  is alternative to the one provided in \cite{rezek2006}. 

The factorized form of a function can be obtained following the use of differential equations, as first proposed by McCoy \cite{mccoy1932a}. 
For any function $g(a\dg, a)$ of the non-commuting operators $[a,a\dg]=c$, the partial derivative can be defined as  \cite{born1926}
 \begin{equation}
c\frac{\partial g}{\partial a\dg} = [a, g],\quad {\rm and } \quad  c\frac{\partial g}{\partial a} = -[a\dg, g].
\end{equation}
Note that $c$ is a constant that will be taken equal to unity at the end, but is useful in the general derivation for the purpose of normalization. 
The expression of the function $g$ in normal ordering (annihilation operators $a$ to the right, creation $a\dg$ to the left) is obtained by integration of a system of partial derivatives. We consider the particular function $\rho = e^{\lambda A\dg A} = \sum_n \frac{\lambda^n}{n!}(A\dg A)^n$, which is quadratic in $a$ and $a\dg$ since, for $A = f_+ a + f_- a\dg$, we have  
\begin{equation}
A\dg A = a\dg a  + f_+^* f_- a^{\dagger 2} + f_+ f_-^* a^2  +c |f_-|^2 \mathbbm{1}.
\end{equation}
To obtain  differential equations, we start from the obvious observation that $(A\dg A)^n A\dg = A\dg (A A\dg)^n$, which gives
\begin{equation} \label{obvious1}
e^{\lambda A\dg A}  A\dg = A\dg e^{\lambda A A\dg}  = e^{c\lambda} A\dg e^{\lambda A\dg A}. 
\end{equation}
Using the relations between the `$A$' and `$a$' operators (\ref{ct}) this readily gives 
\begin{equation}
f_+^* [\rho, a\dg] + f_-^* [\rho, a] = (f_+^* a\dg + f_-^* a) (e^\lambda -1) \rho.
\end{equation}
With the future integration in mind, we write the $a \rho$ term on the r.h.s as $\rho a  +\frac{\partial \rho}{\partial a\dg} $
and obtain the first differential equation
\begin{equation}
cf_+^* \frac{\partial \rho}{\partial a} - cf_-^* e^{c\lambda} \frac{\partial \rho}{\partial a\dg} = f_-^* (e^{c\lambda} -1)\rho a + f_+^* (e^{c\lambda} -1) a\dg \rho.
\end{equation}

A similar equation can be obtained starting from the observation that $A(A\dg A)^n = (A A\dg)^n A$. This gives $[A, \rho] = (e^{c\lambda} -1)\rho A$ and yields to the differential equation
\begin{equation}
- e^{c\lambda} f_-\frac{\partial \rho}{\partial a} + f_+ \frac{\partial \rho}{\partial a\dg} = (e^{c\lambda}-1) f_+ \rho a +f_- (e^{c\lambda} -1) a\dg \rho
\end{equation}
where again, we have chosen  to have terms on the r.h.s in the ordering  $\rho a$ and $a\dg \rho$. 
So we now have the system of differential equation
\begin{equation} \label{sysdiff}
 \begin{cases}
\frac{\partial \rho}{\partial a}  = X \rho a + (e^Y -1) a\dg \rho \\
\frac{\partial \rho}{\partial a\dg} = (e^Y -1) \rho a + X^{*} a\dg \rho  \\
\end{cases}
\end{equation}
with the constants 
\begin{eqnarray}
X &=&\frac{1}{c} \frac{ f_-^* f_+ (e^{2c\lambda}-1) }{|f_+|^2 - |f_-|^2 e^{2 c\lambda}} \\
e^Y - 1  &=&\frac{1}{c} \frac{(e^{c\lambda} -1) (|f_+|^2 + |f_-|^2 e^{c\lambda})}{|f_+|^2 - |f_-|^2 e^{2 c\lambda}} . \equiv y
\end{eqnarray}
It is now easy to verify that  the factorized form (\ref{factorization}) is solution of the system (\ref{sysdiff}). 
So we obtain the following factorized form, in normal ordering
\begin{equation}
\label{eq:reference}
e^{\lambda A\dg A} = e^{\lambda \big(a\dg a  + f_+^* f_- a^{\dagger 2} + f_+ f_-^* a^{2}  + c|f_-|^2 \mathbbm{1}\big)} = K e^{\frac{X^{*}}{2}a{^{\dagger 2}}} e^{\ln(1+y) a\dg a} e^{\frac{X}{2} a^2}.
\end{equation}
For the squeezed thermal state, $f_{+}=\cosh(r_{t})$ and $f_{-}=\sinh(r_{t})e^{i\phi_{t}}$ as follows from (\ref{ct}), which allows to relate directly the squeezing parameters with the factorized form as given in (\ref{eq:facto}).

In order to compute the constant $K$, we further follow the derivation proposed by McCoy \cite{mccoy1932a}. During the factorization, only the commutation relation is important. 
So we choose to replace $a\dg \rightarrow \hat{x}$ and $a\rightarrow c \hat{p} = c \frac{d}{dx}$, with $[\hat{p}, \hat{x}]=c$. 
We look at  how $\lambda A\dg A$ acts on $\mathbbm{1}$, and denote this action   $\lambda A\dg A\{ \mathbbm{1}\}$. With the change of operators, 
$\lambda A\dg A=\lambda\big(f^{*}_{+} x+f^{*}_{-}c\frac{d}{dx}\big)\big(f_{+}c\frac{d}{dx}+f_{-}x  \big)$, so $\lambda A\dg A \{ \mathbbm{1} \}=\lambda\left(|f_{-}|^{2}c+f^{*}_{+}f_{-}x^{2}\right)$. 
The constant term in $(\lambda A\dg A)^{n}$ can be found in applying $n$ times the operator $(\lambda A\dg A)$  on the identity, and is of the form $a_{n}c^{n}$. Let us denote the constant in the first term of the serie $P_{1}(c)$ such that $P_{1}(c)=\sum_{n=0}^{\infty}a_{n}c^{n}/n!$. We denote $P_{2}(c)$ the constant when acting twice $(\lambda A\dg A)$, namely the constant term in  $e^{\lambda A\dg A}\{ \lambda\left(|f_{+}|^2c+f^{*}_{-}f_{+}x^{2}\right)\}$. This yields to $P_{2}(c)=c\frac{\partial P_{1}(c)}{ \partial c}$. We further know from Eq. (\ref{factorization}) that  $P_{2}(c)=K(c)\lambda(c|f_{-}|^{2}+c^{2}f^{*}_{+}f_{-}X)$, which lead to $\frac{\partial K(c)}{\partial c}=K(c)\lambda(c|f_{-}|^{2}+c^{2}f^{*}_{+}f_{-}X)$. At the limit for which the operators commute, $K(c\rightarrow 0)$ tends to unity.  Hence, 
\begin{equation}
\label{eq:kc}
K(c)=\exp\left(\lambda\left(|f_{-}|^{2}c+f^{*}_{+}f_{-}\int_{0}^{c}\zeta X(\zeta)d\zeta\right)\right).
\end{equation}
One can compute the integral 
\begin{align}
\int_{0}^{c}\frac{\zeta}{\zeta}\frac{f^{*}_{+}f_{-}(e^{2\lambda \zeta}-1)}{|f_{+}|^{2}-|f_{-}|^{2}e^{2\lambda \zeta}}d\zeta&=f^{*}_{+}f_{-}\frac{1}{|f_{-}|^{2}}\int_{0}^{c}\left(-1+\frac{|f_{+}|^{2}-|f_{-}|^{2}}{|f_{+}|^{2}-|f_{-}|^{2}e^{2\lambda \zeta}}\right) \nonumber\\
&=-\frac{f^{*}_{+}}{f^{*}_{-}}c+\frac{f^{*}_{+}}{f^{*}_{-}}\ln\left(\frac{|f_{+}|^{2}e^{-2\lambda c}-|f_{-}|^{2}}{|f_{+}|^{2}-|f_{-}|^{2}}\right). 
\end{align}
Inserting this in (\ref{eq:kc}) and using the constant $c=1$, we obtain the normalization constant in the factorized state (\ref{eq:facto}, \ref{factorization}, \ref{eq:reference}) as 
\begin{equation}\label{eq:K}
K=\exp\left(\lambda\left(|f_{-}|^{2}-\frac{f^{*}_{+}}{f^{*}_{-}}\right)\right)\left(\frac{|f_{+}|^{2}e^{-2\lambda }-|f_{-}|^{2}}{|f_{+}|^{2}-|f_{-}|^{2}}\right)^{\frac{f^{*}_{+}}{f^{*}_{-}}}. 
\end{equation}

\section{Effective Hamiltonian in the Unitary case \label{app:unitary}}
We are interested in the evolution of the state $\ket{\Psi_t}$ characterized by the Hamiltonian
\begin{eqnarray}
H&=&- \frac{ \hbar \Delta}{2} \sigma_z +  U_{\rm r,t} H_{\rm int}(t) U_{\rm r,t}\dg.
\end{eqnarray}
Let us first give the explicit form of the interaction Hamiltonian in the rotated frame. 
The atomic part evolves as $e^{\frac{i}{2} (\omega + \Delta) \sigma_z} \sigma_x e^{-\frac{i}{2} (\omega + \Delta) \sigma_z} = e^{- i (\omega+ \Delta)t} \ket{g}\bra{e} + {\rm h.c.}$. The bosonic part is obtained from $e^{i \omega_{0} t a\dg a} a\dg = a\dg e^{i \omega_{0} t (a\dg a + 1)}$ that gives 
$e^{i \omega_{0} t a\dg a}e^{i \eta_l(a\dg + a)}e^{- i \omega_{0} t a\dg a} = e^{i \eta_l (a\dg e^{i \omega_{0} t} + a e^{- i \omega_{0} t})}$. 
Keeping only the terms with the lowest frequency (RWA), we thus have 
\begin{eqnarray}
\label{eq:hintrot}
U_{\rm r,t} H_{\rm int}(t) U_{\rm r,t}\dg &=&\frac{\hbar}{2} \sum_{l=1,2}\Omega_{l}(t)\left(
e^{\frac{i}{2} (\omega + \Delta) \sigma_z} \sigma_x e^{-\frac{i}{2} (\omega + \Delta) \sigma_z}\right) 
\left( e^{i \omega_{0} t a\dg a}e^{i \eta_l(a\dg + a)}e^{- i \omega_{0} t a\dg a}  e^{-i (\Phi_l + \omega_l t)}+ {\rm h.c.} \right)\nonumber
\\
&\approx& \frac{\hbar}{2} \sum_{l=1,2}\Omega_{l}(t)\left(e^{i(\omega_{l}-\omega-\Delta)t}e^{i\Phi_{l}}e^{-i\eta_{l}(a^{\dagger}e^{i\omega_{0} t}+a e^{-i\omega_{0} t})}|g\rangle\langle e| + {\rm h.c.}\right). 
\end{eqnarray}

We are looking for a solution of the wave function as a linear combination of the dressed basis
\begin{equation}
\label{eq:dressed}
|\Psi_{t}\rangle=\sum_n\left(e_{n}(t)|e,n\rangle+g_{n}(t)|g,n\rangle\right).
\end{equation}
The Schrödinger equation gives the excited and ground state populations evolving as 
\begin{subequations}
\begin{align}
\dot{e}_{n}(t)&=i \frac{\Delta}{2} e_{n}(t)-i\sum_{l=1,2}\frac{\Omega_{l}}{2}e^{i(\omega+\Delta-\omega_{l})t}e^{-i\Phi_l} \sum_{n'} \langle n | e^{i\eta_{l}(a\dg_t+a_t)}|n'\rangle g_{n'}(t),\\
 \dot{g}_{n}(t)&={-}i\frac{\Delta}{2}g_n(t){-}i{\sum_{n',l=1,2}}\frac{\Omega_{l}}{2} e^{i(\omega_{l}-\omega-\Delta)t}e^{i\Phi_l}
 \langle n | e^{-i\eta_l(a\dg_t +a_t)}|n'\rangle e_{n'}(t).
\end{align}
\end{subequations}
For large detuning, $|\Delta|\gg|\Omega_{l}|,\omega_{0}$, a state initially  in the electronic ground state mainly remains in this electronic level. The small population of the electronic excited state can be eliminated abiabatically. 
We thus set $\dot{e}_n(t) = 0$. The evolution follows as   
\begin{align} \label{eq:psi1}
\begin{split}
i\hbar\frac{d|\Psi_{t}\rangle}{dt}=&i\hbar\sum_{n}\dot{g}_{n}(t)|g,n\rangle\\
=&\frac{\hbar}{2}\left(\Delta + \frac{\Omega^{2}_{1}+\Omega^{2}_{2}}{\Delta} +\frac{\Omega_1 \Omega_2}{\Delta} (e^{i (\omega_1 - \omega_2)t}e^{i (\Phi_1 - \Phi_2)} e^{i (\eta_2 - \eta_1) (a\dg_t+ a_t)} + {\rm h.c.}) \right) \ket{g}\braket{g}{\Psi_t}\\
=&H_{\rm eff} \ket{\Psi_t},
\end{split}
\end{align}
where we have defined the effective Hamiltonian 
\begin{align}
H_{\rm eff}&=\frac{\hbar}{2}\left(\Delta + \frac{\Omega^{2}_{1}+\Omega^{2}_{2}}{\Delta}\right)|g\rangle\langle g|+\hbar\frac{\Omega_{1}\Omega_{2}}{2\Delta}\left(e^{i(\omega_{1}-\omega_{2})t}e^{i(\Phi_{1}-\Phi_{2})} e^{i(\eta_{2}-\eta_{1})(a^{\dagger}_t+a_t)}+{\rm h.c.}\right)|g\rangle\langle g|.
\end{align}
We then use Glauber formula to write $e^{i (\eta_2 - \eta_1) (a_t\dg+a_t )} = e^{i (\eta_2 - \eta_1) a_t\dg}e^{i (\eta_2 - \eta_1) a_t}e^{- (\eta_2 - \eta_1)^2/2}$ and expand the exponentials in series to keep only the first resonant term. For $\omega_1 - \omega_2 = 2 \omega_{0}$, this leads, in leading order of $(\eta_2 - \eta_1)$, to 
\begin{align}
\label{eq:hefftilde}
H_{\rm eff}&\approx \frac{\hbar}{2}\left(\Delta + \frac{\Omega^{2}_{1}+\Omega^{2}_{2}}{\Delta}\right)|g\rangle\langle g|
+ \frac{\hbar}{4}(\eta_{2}-\eta_{1})^{2}\frac{\Omega_{1}\Omega_{2}}{\Delta}\left(e^{i(\Phi_{1}-\Phi_{2})} a^{2}+{\rm h.c.}\right)|g\rangle\langle g|.
\end{align}
This corresponds, up to the Lamb-shift term that we neglect, to the effective Hamiltonian given in Eq. (\ref{eq:heff}) of the main text.

\section{Dynamics for an ion in a stochastically shaken trap and driven with two-photon Raman interaction \label{app:trap}}
We are interested in the evolution of the state $\ket{\Psi_t} = U_{\rm r,t} \ket{\psi_t}$  characterized by the stochastic Hamiltonian
\begin{eqnarray}
H_{st}&=&U_{r,t}h_{\rm st}U^{\dagger}_{r,t}=- \frac{ \hbar \Delta}{2} \sigma_z +  U_{\rm r,t} H_{\rm int}(t) U_{\rm r,t}\dg + \hbar \sqrt{2\gamma_{t}}\xi_{t}U_{{\rm r},t}\hat{x}U_{{\rm r},t}\dg\otimes|g \rangle \langle g |. 
\end{eqnarray}

The state vector $|\Psi_{t}\rangle$ follows a stochastic evolution. 
Over a small time increment, it evolves as 
$|\Psi_{t+dt}\rangle= e^{- \frac{i}{\hbar} H_{\rm st} dt} \ket{\Psi_t} = \exp\Big({-}\frac{i}{\hbar}\big(H dt-i \sqrt{2\gamma_{t}}\hat{x}_t d W_{t}\otimes |g\rangle \langle g|\Big)|\Psi_{t}\rangle $, where $dW_t = \xi_t dt$ represents the differential Wiener increment. The later verifies Itô rules for stochastic calculus. In particular, $(dW_{t})^{2}=dt$ and $dW_{t}dt=0$ \cite{jacobs2006}, such that a Taylor expansion of the exponential yields
$d|\Psi_{t}\rangle=\Big(-\frac{i}{\hbar}H dt- \Big(i \sqrt{2\gamma_{t}}\hat{x}_t dW_{t}+\gamma_{t}\hat{x}^{2}_t dt \Big) \otimes |g\rangle \langle g|\Big)|\Psi_{t}\rangle.$
Starting back from (\ref{eq:hintrot}), we look for the solution as a linear combination of the dressed basis (\ref{eq:dressed}). The evolution of this state over a small increment of time $dt$ gives the excited and ground state populations evolving as 
\begin{subequations}
\begin{align}
\dot{e}_{n}(t)&=i \frac{\Delta}{2} e_{n}(t)-i\sum_{l=1,2}\frac{\Omega_{l}}{2}e^{i(\omega+\Delta-\omega_{l})t}e^{-i\Phi_l} \sum_{n'} \langle n | e^{i\eta_{l}(a\dg_t+a_t)}|n'\rangle g_{n'}(t),\\
 \dot{g}_{n}(t)&={-}i\frac{\Delta}{2}g_n(t){-}i{\sum_{n',l=1,2}}\frac{\Omega_{l}}{2} e^{i(\omega_{l}-\omega-\Delta)t}e^{i\Phi_l}
 \langle n | e^{-i\eta_l(a\dg_t +a_t)}|n'\rangle e_{n'}(t)  {-} \sum_{n'}\langle n |  \Big(i \sqrt{2\gamma_{t}}\hat{x}_t \frac{dW_{t}}{dt} {+} \gamma_{t} \hat{x}^{2}_t\Big) |n'\rangle g_{n'}(t).
\end{align}
\end{subequations}
For large detuning, $|\Delta|\gg|\Omega_{l}|,\omega_{0}$, a state initially  in the electronic ground state mainly remains in this electronic level. The small population of the electronic excited state can be eliminated abiabatically. 
We thus set $\dot{e}_n(t) = 0$, and the evolution follows as   
\begin{align} \label{eq:psi1}
\begin{split}
i\hbar\frac{d|\Psi_{t}\rangle}{dt}=&i\hbar\sum_{n}\dot{g}_{n}(t)|g,n\rangle\\
=&\frac{\hbar}{2}\left(\Delta + \frac{\Omega^{2}_{1}+\Omega^{2}_{2}}{\Delta} +\frac{\Omega_1 \Omega_2}{\Delta} (e^{i (\omega_1 - \omega_2)t}e^{i (\Phi_1 - \Phi_2)} e^{i (\eta_2 - \eta_1) (a\dg_t+ a_t)} + {\rm h.c.}) \right) \ket{g}\braket{g}{\Psi_t}\\
&- i\hbar  \Big(i\sqrt{2\gamma_{t}}\hat{x}_t \frac{dW_{t}}{dt} {+} \gamma_{t} \hat{x}^{2}_t\Big) \sum_{n'}|g, n'\rangle \langle g,n' |\Psi_{t}\rangle, \\
=&H_{\rm eff} \ket{\Psi_t} -i\hbar  \Big(i\sqrt{2\gamma_{t}}\hat{x}_t \frac{dW_{t}}{dt}+\gamma_{t} \hat{x}^{2}_t\Big) |g\rangle \langle g |\Psi_{t}\rangle,
\end{split}
\end{align}
with the  effective Hamiltonian equal to Eq. (\ref{eq:hefftilde}).
After the adiabatic elimination, the increment reads $ d \ket{\Psi_t} =\left( - \frac{i}{\hbar}H_{\rm eff}(t) dt - (i\sqrt{2 \gamma_t} dW_t \hat{x}_t +  \gamma_t  dt \hat{x}^2_t)  \ket{g} \bra{g} \right) \ket{\Psi_t}$.

It is now easy to characterize the evolution of the density matrix $\rho_{\rm st} = \ket{\Psi_{t}}\bra{\Psi_{t}}$. The Leibnitz chain rule that, in the It\^o calculus, 
  generalizes to $d(AB) = (A+ dA) (B+dB) - AB = (dA) B + A (dB) + dA dB $  \cite{Adler2003a}, yields  
\begin{equation}
d \rho_{\rm st}=- \frac{i}{\hbar} [H_{\rm eff}(t), \rho_{\rm st}] dt - i \sqrt{2\gamma_t} [\hat{x}_t{\otimes} |g\rangle \langle g|,\rho_{\rm st} ] dW_t -\gamma_t 
[\hat{x}_t \otimes |g\rangle \langle g|,[\hat{x}_t\otimes |g\rangle \langle g|,\rho_{\rm st} ]]dt,
\end{equation}
which preserves the norm at the level of each individual stochastic realization. 
The density matrix of interest here in the ensemble one, obtained from averaging over the realizations of the noise and  denoted $\rho_t= \la \rho_{\rm st}\ra$. 
Since the average of any function $F_t$ of the stochastic process vanishes, $\la F_t dW_t\ra=0$ \cite{Adler2003a}, we find that the evolution of the ensemble density matrix $\rho_t$ is dictated by the master equation (\ref{eq:mastereqrot}) given in the main text. 

We now solve this equation and find the control parameters for which the squeezed thermal state (\ref{eq:solution}) is a solution. 
To do so, we look at $\frac{d \rho}{dt} \rho^{-1}$ and  use the factorized form of the squeezed thermal state that allows recasting all needed terms of the master equation (\ref{eq:rhoirho}) in the form $e^{A} B e^{-A}$ for all elements $\{A,B\}$ in the $\mathcal{B}\equiv\{a^{\dagger2},a^{2},\mathbbm{1},a^{\dagger}a,a^{\dagger},a\}$ basis.  
We denote these terms with the adjoint operator $\mathbbm{A}$ of an operator $A$,  defined by recurrence from $\mathbbm{A}_A^n B = [A, \mathbbm{A}_A^{n-1}]$ with $\mathbbm{A}_A^1 B = [A,B]$ and $\mathbbm{A}^0_A = 1$. 
The BCH formula then becomes 
\begin{eqnarray}
e^{A} B e^{-A} &=& B + [A,B] + \frac{1}{2!} [A, [A,B]] + \frac{1}{3!} [A,[A,[A,B]]] + \dots \nonumber \\
&=& \sum_{n=0}^\infty \frac{\mathbbm{A}_A^n}{n!} B = e^{\mathbbm{A}_A} B.
\end{eqnarray}

 We then explicit the transformation for each element 
of the basis. The action of the adjoint is thus a  linear transformation that can be represented in matrix form  in the basis $\mathcal{B}$, the needed terms being explicitly 
{\small
\[ \begin{split}
  \quad \quad e^{\mathbbm{A}_{J^{*}a^{\dagger2}}}
  =  &\begin{pmatrix}
 1 & 4J^{*2}& 0 & -2J^{*} & 0 &0 \\
 0 & 1 & 0 & 0 & 0 &0\\
 0& -2J^{*} & 1 & 0 & 0 &0 \\
 0 & -4J^{*} & 0 &1 &0 &0 \\
 0 & 0 & 0 & 0 & 1 & -2J^{*}\\
  0 & 0 & 0 & 0 & 0 & 1 \\
  \end{pmatrix},
 \end{split}
 \qquad
 \begin{split}
  \quad \quad e^{\mathbbm{A}_{Ja^{2}}}
  = &\begin{pmatrix}
1 & 0 &0 &0 &0 &0\\
4J^{2} & 1 & 0 & 2J & 0 &0 \\
2J& 0& 1 & 0 &0 &0 \\
4J& 0 &0 &1 & 0 &0 \\
0 &0 &0 &0 & 1 &0 \\
0 &0 &0 &0 & 2J & 1\\
  \end{pmatrix},
  \end{split}
\]}
and
\begin{equation}
e^{\mathbbm{A}_{-B a^{\dagger}a}}= \rm{diag}\left(e^{-2 B},e^{2 B},1,1,e^{- B},e^{B}\right).
\end{equation}
Equation (\ref{eq:rhoirho}) then follows, in matrix representation in the $\mathcal{B}$ basis, as 
\begin{align}
\begin{split}
\frac{d\rho}{dt}\rho^{-1}&=\begin{pmatrix}
\dot{J}^{*}+2\dot{B}J^{*}+4e^{2B}\dot{J}(J^{*})^{2}\\
e^{2B}\dot{J}\\
\frac{d}{dt}\left(\frac{1}{Z_{t}}\right)Z_{t}-2e^{2B}\dot{J}J^{*}\\
-\dot{B}-4e^{2B}\dot{J}J^{*}\\
0\\
0
\end{pmatrix}=
\begin{pmatrix}
-i\alpha_{R}-\alpha_{I}\\
-i\alpha_{R}+\alpha_{I}\\
-2\kappa\\
-2\kappa\\
0\\
0
\end{pmatrix}
+\rho\begin{pmatrix}
(i\alpha_{R}+\alpha_{I})\\
(i\alpha_{R}-\alpha_{I})\\
0\\
-2\kappa\\
0\\
0
\end{pmatrix}\rho^{-1}
+2\kappa a\rho a^{\dagger}\rho^{-1}
+2\kappa a^{\dagger}\rho a \rho^{-1}.\\
&=\begin{pmatrix}
-i\alpha_{R}-\alpha_{I}\\
-i\alpha_{R}+\alpha_{I}\\
-2\kappa\\
-2\kappa\\
0\\
0
\end{pmatrix}
+e^{\mathbbm{A}_{J^{*}a^{\dagger 2}}}e^{\mathbbm{A}_{-Ba^{\dagger}a}}e^{\mathbbm{A}_{Ja^{2}}}\begin{pmatrix}
(i\alpha_{R}+\alpha_{I})\\
(i\alpha_{R}-\alpha_{I})\\
0\\
-2\kappa\\
0\\
0
\end{pmatrix}
+2\kappa\begin{pmatrix}
0\\
2e^{B}J_t\\
e^{-B}-4e^{B}J_tJ^{*}_t\\
e^{-B}-4e^{B}J_tJ^{*}_t\\
0\\
0
\end{pmatrix}
+2\kappa\begin{pmatrix}
-2e^{B}J^{*}_t\\
0\\
0\\
e^{B}\\
0\\
0
\end{pmatrix} \label{eq:leftside}
.
\end{split}
\end{align}
Finally, 
\begin{frame}
\tiny
\setlength{\arraycolsep}{2.5pt} 
\medmuskip = 1mu 
\begin{align}
\frac{d\rho}{dt}\rho^{-1}=
\resizebox{.8\hsize}{!}{$\begin{pmatrix}
&i\alpha_{R}\left(-1+4e^{2B}J^{*2}_t+e^{-2B}(1-4e^{2B}J_tJ^{*}_t)^{2}\right)+\alpha_{I}\left(-1-4e^{2B}J^{*2}_t +e^{-2B}(1-4e^{2B}J_t J^{*}_t)\right)+\kappa_t \left(-4e^{B}J^{*}_t-4J^{*}_t(-1+4e^{2B}J_t J^{*}_t)\right)\\
&i\alpha_{R}\left(-1+e^{2B}(1+4J^{2}_t)\right)+\alpha_{I}\left(1+e^{2B}(-1+4J^{2}_t)\right)-4e^{B}(-1+e^{B})J_t\kappa_t\\
&2i\alpha_{R}\left(J_t-e^{2B}(1+4J^{2}_t)J^{*}_t\right)+2\alpha_{I}\left(J_t+e^{2B}(1-4J^{2}_t)J^{*}_t\right)+\kappa_t\left(2(-1+e^{-B})-8e^{B}|J_t|^{2}+8e^{2B}|J_t|^{2}\right)\\
&4i\alpha_{R}\left(J_t-e^{2B}(1+4J^{2}_t)J^{*}_t\right)+4\alpha_{I}\left(J_t+e^{2B}(1-4J^{2}_t)J^{*}_t\right)+\kappa_t\left(4(-1+\cosh(B))-8e^{B}|J_t|^{2}+16e^{2B}|J_{t}|^{2}\right)
\end{pmatrix}$}.\label{eq:rightside}
\end{align}
\end{frame}

By linear combination of equations of the system (\ref{eq:rightside}), we can  identify the evolution parameters of the squeezed thermal state and obtain the coupled differential equations 
\begin{subequations}
\label{eq:resultdyn}
\begin{align}
\label{eq:resultdyn1}
\dot{J_{t}}&=-4e^{-B}\left(-1+e^{B}\right)J_{t}\kappa+i(-e^{-2B}+(1+4J_{t}^{2}))\alpha_{R}+\left(e^{-2B}+(-1+4J^{2}_{t})\right)\alpha_{I},\\
\label{eq:resultdyn2}
\dot{B}&=-4\left(\kappa\left(-1+\cosh(B)+2e^{B}|J_{t}|^{2}\right)+i\alpha_{R}(J_{t}-J^{*}_{t})+\alpha_{I}(J_{t}+J^{*}_{t})\right)
\end{align}
\end{subequations}
that provide the control parameters as function of the state characteristics, as given in matrix form in the main text \eqref{eq:dyneq1}.

\section{Two-photon Raman interaction and stochastically driven Jaynes-Cummings Hamiltonian \label{sec:laser}}
We now consider to generate the dissipator through a stochastic laser field rather than through shaking the trap. This leads to a  Jaynes-Cummings (JC) Hamiltonian \cite{Meystre1982,PhysRevA.37.3175} in its stochastic form \cite{Lawande1994,Qiu2017}.  Note that the effect of dissipation  in the JC model has been considered \cite{Scala2007,Scala2007bis,Kim1994}, mainly focusing on the influence over the populations. 

\begin{figure}
\begin{center}
\includegraphics[width=0.35\columnwidth]{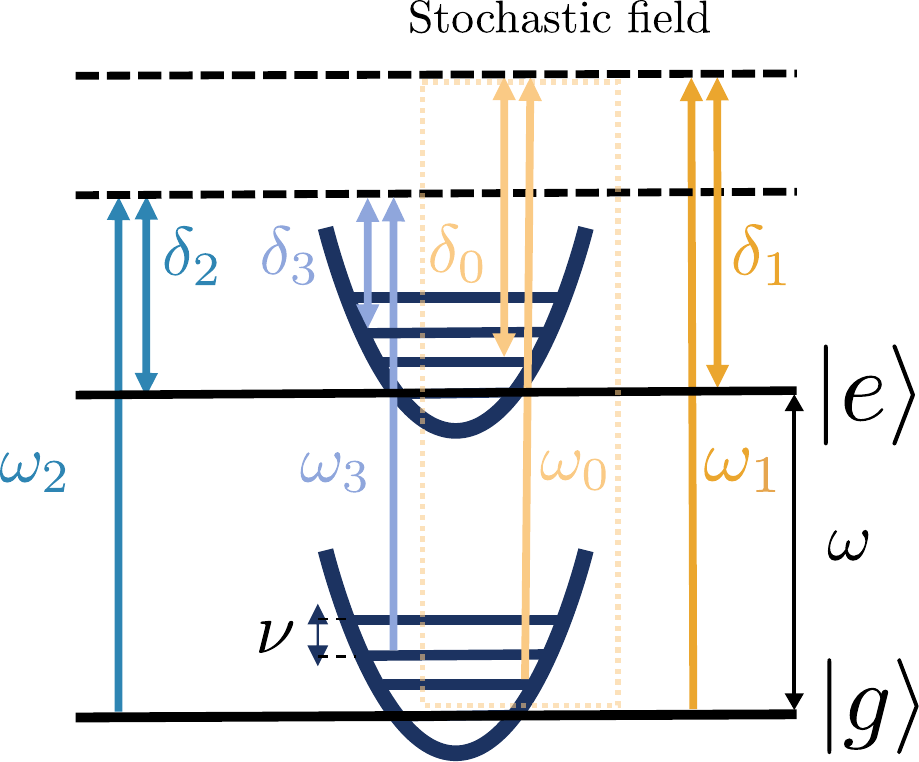}
\end{center}
\captionof{figure}{{\bf Experimental setup}: 2-photon Raman interaction is generated by the (blue) laser pair with $\omega_2 - \omega_3 = 2\nu$, while dephasing is generated with the (red) laser pair, $\omega_1 - \omega_0 = \nu$, one amplitude being taken as stochastic.  \label{fig:setup2}}
\end{figure}

The set-up is similar to the one presented in Sec. \ref{sec:exp}, but with 
  two additional beams  used to engineer the dissipator (see Fig. \ref{fig:setup2} for an  illustration).
The interaction Hamiltonian resulting from the applied  laser fields now reads  \cite{leibfried2003}
\begin{equation}
\label{eq:Hint}
H_{\rm int}(t) = {\sum_{l=\{0,\dots,3\}}}  \frac{\hbar}{2}\Omega_l \sigma_x \left( e^{i (k_l \hat{x} - \omega_l t - \Phi_l)} + {\rm h.c.}\right),  
\end{equation}
where the Rabi frequency $\Omega_0$ will be taken as stochastic $\Omega_0^{\rm st}$. 
We aim at preparing a squeezed thermal state on the vibrational levels of the system, $H_m = \hbar \nu (a\dg a + 1/2)$,  with total Hamiltonian
\begin{equation}
h_{\rm tot}(t)=H_{\rm a}+H_{\rm m}+H_{\rm int}(t).
\end{equation}
starting from an initial vibrational state that is thermal. As discussed above, this will be done by reverse engineering of the master equation to allow both squeezing and thermalization. 
We proceed as before and look at  the evolution of the rotated vector $\ket{\Psi_t}\equiv U_{\rm r,t} |\Psi_{t}\rangle$. The unitary $U_{\rm r,t}\equiv e^{\frac{i}{\hbar}H_{r}t}$ is defined from the rotation Hamiltonian $H_{\rm r}=H_{\rm a}+H_{\rm m}+\frac{\hbar\tilde{\Delta}}{2}\sigma_{z}$, the average detuning now being $\tilde{\Delta} = \sum_{l=\{0,\dots,3\}} \delta_l/4$. 
The rotated state evolves as  $\ket{\dot{\Psi}_{t+dt}} = e^{-\frac{i}{\hbar}  H_{\rm tot} dt}\ket{\Psi_t}$ with
 $H_{\rm tot}\equiv U_{\rm r,t} h_{\rm tot} U_{\rm r,t}\dg + i \hbar \dot{U}_{\rm r,t} U_{\rm r,t}\dg=- \frac{ \hbar \tilde{\Delta}}{2} \sigma_z +  U_{\rm r,t} H_{\rm int}(t) U_{\rm r,t}\dg$. The interaction Hamiltonian in the rotated frame, after the RWA, reads \begin{eqnarray}
U_{\rm r,t} H_{\rm int}(t) U_{\rm r,t}\dg &=&\frac{\hbar}{2} \sum_{l}\Omega_{l}(t)\left(
e^{\frac{i}{2} (\omega + \tilde{\Delta}) \sigma_z} \sigma_x e^{-\frac{i}{2} (\omega + \tilde{\Delta}) \sigma_z}\right) 
\left( e^{i \omega_{0} t a\dg a}e^{i \eta_l(a\dg + a)}e^{- i \omega_{0} t a\dg a}  e^{-i (\Phi_l + \omega_l t)}+ {\rm h.c.} \right)  \nonumber \\
&\approx& \frac{\hbar}{2} \sum_{l}\Omega_{l}(t)\left(\hat{h}_l |g\rangle\langle e| + {\rm h.c.}\right), 
\end{eqnarray}
where we have defined $\hat{h}_l \equiv e^{- i (\omega - \tilde{\Delta} - \omega_l)t}e^{i \Phi_l}e^{- i \eta_l (a\dg_t + a_t)}$.

The open dynamics is generated using a white noise on top of the `0' laser's amplitude, namely taking $\Omega^{\rm st}_0 \rightarrow \sqrt{\Omega_0}\xi_t$. 
 It is then  convenient to split the total Hamiltonian  into its deterministic  and stochastic  contributions $H_{\rm tot} = H_{\rm det}+ \xi_t H_0$, defined as 
\begin{subequations}
\begin{align}
H_{\rm det} &= - \hbar \frac{\tilde{\Delta}}{2} \sigma_z + \frac{\hbar}{2} \sum_{l=\{1,2,3\}} \Omega_l (\hat{h}_l \ket{g}\bra{e} + \hat{h}_l\dg \ket{e}\bra{g}) \\
\xi_t H_0&=\xi_t \frac{\hbar}{2}\sqrt{\Omega_0}(\hat{h}_0 \ket{g}\bra{e} + \hat{h}_0\dg \ket{e}\bra{g}) . 
\end{align}
\end{subequations}

We look for a solution of the wave function as 
$
|\Psi_{t}\rangle=\sum_{n=0}^\infty\left(e_{n}(t)|e,n\rangle+g_{n}(t)|g,n\rangle\right).
$
The evolution of this state over a small increment of time $dt$ reads $d\ket{\Psi_t} =   -\frac{i}{\hbar} (H_{\rm det}dt + H_{0} dW_t)  -\frac{1}{2\hbar^2}H_{0}^2 dt$. This yields the coefficients evolving as 
\begin{subequations}
\begin{align}
\dot{e}_{n}(t)&=i \frac{\tilde{\Delta}}{2} e_{n}(t)-\frac{i}{2} \sum_{n'} \left( \sum_{l\neq 0}  \Omega_{l} \bra{n}\hat{h}_l\dg \ket{n'}  + 
\sqrt{\Omega_{0}}\xi_t \bra{n}\hat{h}_0\dg \ket{n'}  \right) g_{n'}(t) - \frac{1}{8} \Omega_0\bra{n} \hat{h}_0\dg \hat{h}_0 \ket{n'} e_{n'}(t)\\
\dot{g}_{n}(t)&=-i \frac{\tilde{\Delta}}{2} g_{n}(t)-\frac{i}{2} \sum_{n'} \left( \sum_{l\neq 0}  \Omega_{l} \bra{n}\hat{h}_l\dg \ket{n'}  + 
\sqrt{\Omega_{0}}\xi_t \bra{n}\hat{h}_0\dg \ket{n'}  \right) e_{n'}(t) - \frac{1}{8} \Omega_0\bra{n} \hat{h}_0\dg \hat{h}_0 \ket{n'} g_{n'}(t)
\end{align}
\end{subequations}

For large detuning, $|\tilde{\Delta}|\gg|\Omega_{l}|,\nu$, a state initially  in the electronic ground state mainly remains in this electronic level. 
The small population of the electronic excited state can be eliminated abiabatically. 
We thus set $\dot{e}_n(t) = 0$, and the evolution follows as   (assuming $\frac{\Omega_0}{\tilde{\Delta}}\ll1$)
\begin{eqnarray}
i\hbar\frac{d|\Psi_{t}\rangle}{dt}
&=&\frac{\hbar}{2}\left(\tilde{\Delta} + \frac{\xi_t \sqrt{\Omega_0}}{\tilde{\Delta}} - i \frac{\Omega_0}{4} 
+  \sum_{l\neq 0} 
 \frac{\Omega_l}{\tilde{\Delta}} \Big( \sum_{l'\neq 0} \Omega_{l'} \hat{h}_0 \hat{h}_{l'}\dg + \Omega_0 \xi_t (\hat{h}_l \hat{h}_0\dg + \hat{h}_0 \hat{h}_{l}\dg) \Big) \right) \ket{g}\braket{g}{\Psi_t}.  
\end{eqnarray}
We then split the term $e^{-i (\eta_{l} - \eta_{l'}) (a_t\dg+a_t )}$ and expand the exponentials in series to keep only the first resonant term. 
Choosing $\omega_2 - \omega_3 = 2 \nu$, the first resonant term brings a quadratic contribution of the form $a^2 e^{i (\Phi_2 - \Phi_3)}$; and $\omega_1 - \omega_0 =  \nu$ gives the slowest oscillating term as linear, $a e^{i (\Phi_1 - \Phi_0)}$. 
Thus, the resonant contributions are between the pairs of lasers, and read, in leading order of $(\eta_l - \eta_{l'})$,  
\begin{eqnarray}
\hat{h}_l \hat{h}_{l'\neq l}\dg  &=& e^{i (\omega_l - \omega_{l'})t} e^{i (\Phi_l - \Phi_{l'})}e^{-i (\eta_l - \eta_{l'}) (a_t\dg+a_t )} \nonumber \\
&=& e^{i (\omega_l - \omega_{l'})t} e^{i (\Phi_l - \Phi_{l'})} \sum_{j,j'}(-i)^{j+j'}\frac{(\eta_j - \eta_{j'})^{j+j'}}{j!j'!} a^{\dagger j} a^{j'} e^{i \omega_{0} t (j- j')} e^{- (\eta_{l'} - \eta_l)^2/2} \nonumber \\
&\approx& \delta_{l,1} \delta_{l',2} \frac{(-i)^2}{2!}(\eta_2 - \eta_3)^2 \left( a^2 e^{i (\Phi_2 - \Phi_3)} + {\rm h.c.}\right)  
-i \delta_{l,1} \delta_{l',0}(\eta_1 - \eta_0)(a e^{i (\Phi_1 - \Phi_0)} - {\rm h.c.}).
\end{eqnarray}
The evolution of the wave function then becomes 
\begin{eqnarray}
i\hbar\frac{d|\Psi_{t}\rangle}{dt}&=&\frac{\hbar}{2}\left(\tilde{\Delta} + \frac{\sum_{l\neq 0}\Omega_l^2 + \xi_t \Omega_{0}}{\tilde{\Delta}} -\frac{1}{2} \frac{\Omega_2 \Omega_{3}}{\tilde{\Delta}} (\eta_2 - \eta_3)^2 \left( a^2 e^{i (\Phi_2 - \Phi_3)} + {\rm h.c.}\right)\right) \ket{g}\braket{g}{\Psi_t}  \\
&& - i\frac{\hbar}{2}\left( \frac{\Omega_{0}}{4} + \frac{\Omega_1 \sqrt{\Omega_0}}{\tilde{\Delta}} \xi_t (\eta_1 - \eta_0)(a e^{i (\Phi_1 - \Phi_0)} - {\rm h.c.}) \right)  \ket{g}\braket{g}{\Psi_t}.  \nonumber
\end{eqnarray}
We can thus define an effective Hamiltonian 
 $H_{\rm sq} \equiv \left( \alpha_t a^2 + {\rm h.c.}\right) \ket{g}\bra{g}$, where  $\alpha_t=- \frac{\Omega_2 \Omega_{3}}{4\tilde{\Delta}} (\eta_2 - \eta_3)^2 e^{i (\Phi_2 - \Phi_3)} $, 
  and a dissipator 
$D_a =  \hbar \frac{\Omega_1 \sqrt{\Omega_0}}{2\tilde{\Delta}}  (\eta_1 - \eta_0)(a e^{i (\Phi_1 - \Phi_0)} -{\rm h.c.}) \ket{g}\bra{g} $ and obtain the  compact expression (neglecting the Lamb shift)
\begin{equation}
i\hbar\frac{d|\Psi_{t}\rangle}{dt} = H_{\rm sq} \ket{\Psi_t} - i (\hbar \frac{\Omega_0}{8}\ket{g}\bra{g} + \xi_t D_a )\ket{\Psi_t}. 
\end{equation}
 Using the previously defined Leibnitz chain rule, we obtain the master equation for the noise-average density matrix  
 \begin{subequations}
\begin{align}
\label{eq:mastereq}
\frac{d \rho_t }{dt} &= - \frac{i}{\hbar} [H_{\rm sq}, \rho_t] -  \hbar\frac{\Omega_0}{8}\{\ket{g}\bra{g} , \rho_t\big\} + \frac{1}{4}D_a  \rho_t  D_a\dg,\\
&=-i [\alpha_t a^{2}+ \alpha_t^* a^{\dagger 2},\rho_t]-  \hbar\frac{\Omega_0}{8} (\ket{g}\bra{g}  \rho_t +  \rho_t  \ket{g}\bra{g}){+}\frac{\kappa_t}{4}\left(a+a\dg \right)\rho_t \left(a+a\dg \right). \label{eq:submastereq}
\end{align}
\end{subequations} 
In the second line, we have applied the RWA, set $\Phi_1-\Phi_0= \pi/2$, and defined $\kappa_{t}= \big(  \hbar \frac{\Omega_1 \sqrt{\Omega_0}}{2\tilde{\Delta}} (\eta_1 - \eta_0) \big)^{2}$ to express the dissipator as $D_a =  (i\sqrt{\kappa_t} a \ket{g}\bra{g} -{\rm h.c.})$.

We next solve the dynamics to find the dynamical  control parameters $\{\alpha_t, \kappa_t\}$ for which the  squeezed thermal state 
$|g\rangle\langle g| \otimes \frac{K_t }{Z_{t}}e^{J^{*}_t a^{\dagger 2}}e^{-B_t a^{\dagger}a}e^{J_t a^{2}}$ is solution of (\ref{eq:submastereq}). 
Proceeding similarly to the other setup,  the master equation (\ref{eq:submastereq}) is rewritten in the basis $\mathcal{B}$ and now reads 
\begin{align}
\label{eq:eqdyntwo}
\begin{split}
\frac{d\rho_t}{dt}\rho_t^{-1}&=\begin{pmatrix}
-i\alpha_{R}-\alpha_{I}\\
-i\alpha_{R}+\alpha_{I}\\
0\\
0\\
0\\
0
\end{pmatrix}
+e^{\mathbbm{A}_{J^{*}a^{\dagger 2}}}e^{\mathbbm{A}_{-Ba^{\dagger}a}}e^{\mathbbm{A}_{Ja^{2}}}\begin{pmatrix}
(i\alpha_{R}+\alpha_{I})\\
(i\alpha_{R}-\alpha_{I})\\
0\\
0\\
0\\
0
\end{pmatrix}
\\
&+\frac{1}{4}\kappa_t \left(a+a\dg \right)e^{\mathbbm{A}_{J^{*}a^{\dagger 2}}}e^{\mathbbm{A}_{-Ba^{\dagger}a}}e^{\mathbbm{A}_{Ja^{2}}} \left(a+a\dg \right)-\hbar\frac{\Omega_0}{4}\mathbbm{1}.\\
\end{split}
\end{align}

By linear combination of the equations in the system (\ref{eq:eqdyntwo}), the control parameters  are found as solutions of \begin{subequations}
\label{eq:resultdyntwo}
\begin{align}
\label{eq:resultdyn1}
\dot{J}_{t}&=\frac{1}{4}e^{-B}(1+2J_t)\kappa_t+i\alpha_R(-e^{-2B}+1+4J_t^2)+\alpha_I(e^{-2B}-1+4J_t^2)\\
\label{eq:resultdyn2}
\dot{B}&=-\frac{1}{4}\left(e^{-B}+e^{B}(1+2J_t)(1+2J^{*}_t)\right)\kappa_t-4i\alpha_R(J_t-J^{*}_t)-4\alpha_I(J_t+J^{*}_t). 
\end{align}
\end{subequations}

So this dynamics creates the squeezed thermal state (\ref{eq:solution})
provided that the control parameters fulfill 
\begin{align}
\label{eq:dyneq}
\begin{pmatrix}
\kappa\\
\alpha_{R}\\
\alpha_{I}\\
\end{pmatrix}
=M_{t}^{-1}
\begin{pmatrix}
\dot{J}_{R}\\
\dot{J}_{I}\\
\dot{B}
\end{pmatrix}, 
\end{align}
with the matrix now reading 
\begin{frame}
\footnotesize
\setlength{\arraycolsep}{2.5pt} 
\medmuskip = 1mu 
\begin{align}
&M_{t}=
&\begin{pmatrix}
-\frac{1}{4}e^{-B}(1+2J_{R})&-8J_{I}J_{R}&4(J^{2}_{R}-J^{2}_{I})+(e^{-2B}-1)\\
\frac{1}{2}e^{-B}J_{I}&4(J^{2}_{R}-J^{2}_{I})+ (1-e^{-2B})&8J_{R}J_{I}\\
-\left(\frac{1}{2} \cosh B + e^{B} (J_{R}^2 + J_I^2+J_{R})\right)&8J_{I}&-8J_{R}
\end{pmatrix}.
\end{align}
\end{frame}

\begin{figure}
\centering
\includegraphics[width = 1\columnwidth]{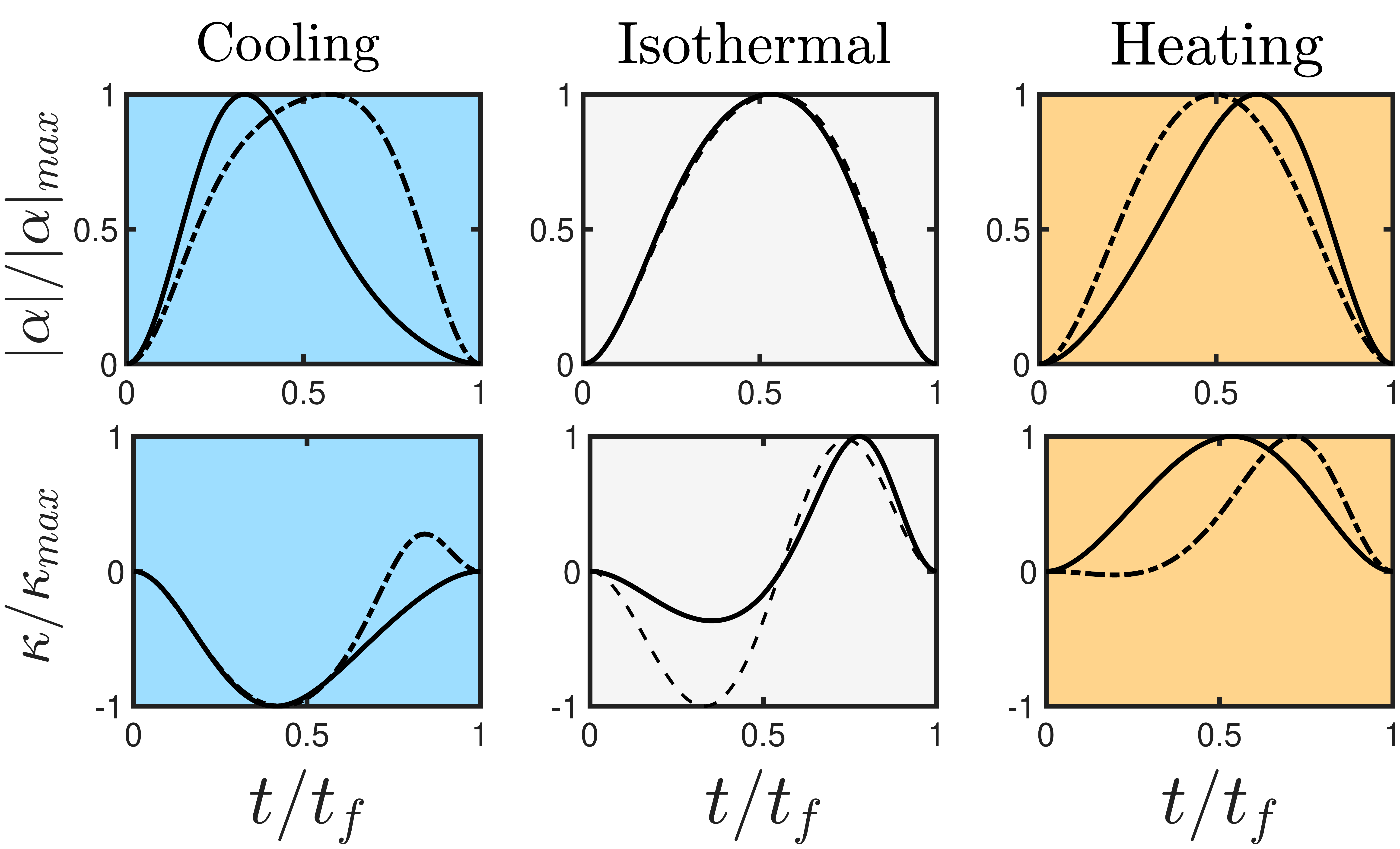}
\captionof{figure}{\textbf{Control parameters}: relative laser amplitude (top) and dephasing strength  (bottom)  for (a) cooling ($\lambda_f=-2$), (b) isothermal ($\lambda_f=\lambda_i$), and (c) heating ($\lambda_f=-0.5$) processes. The initial state is isotropic  $r_{i}=\phi_{i}=0$ at $\lambda_i=-1$. The final state ($t_f=1$) is a thermal state with 
 no squeezing  $r_{f}=\phi_{f}=0$ (plain lines); squeezing at $(r_{f}=1, \phi_f =0)$  (dash-dotted lines), or  squeezing at $r_{f}=1$ and angle $\phi_{f}=\frac{\pi}{4}$ (dashed lines). 
 \label{fig:control4laser}}
\end{figure}

\begin{figure}
\centering 
\includegraphics[width=1\columnwidth]{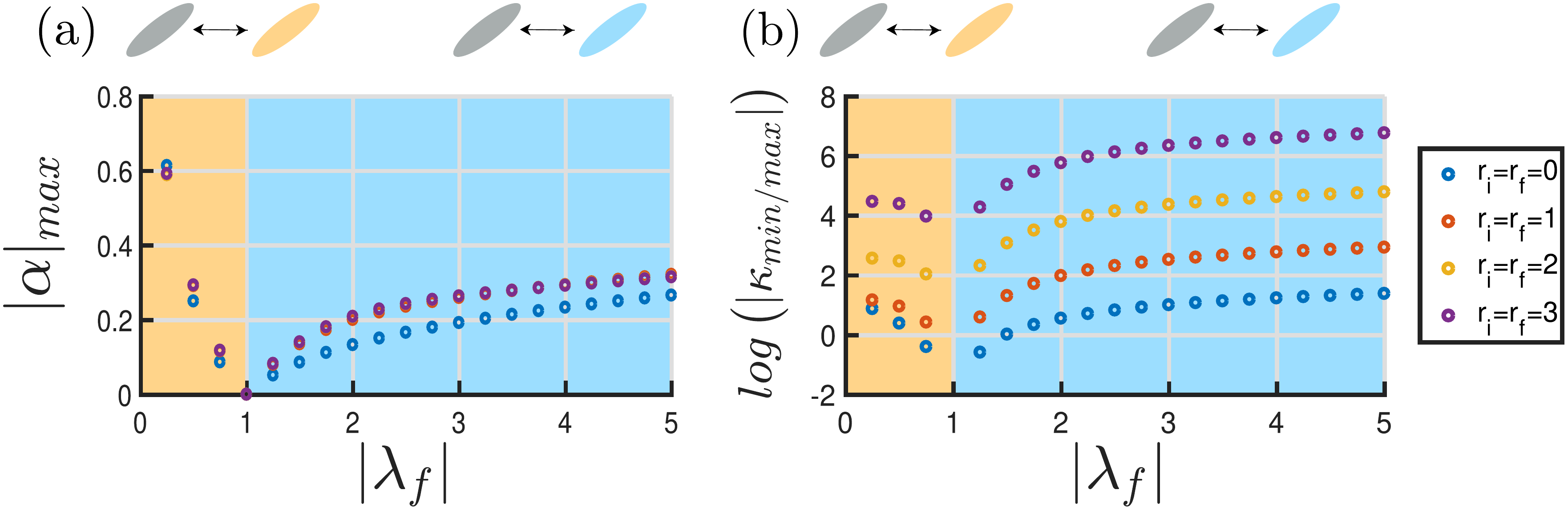}
\captionof{figure}{{\bf Influence of temperature on the control maxima}:  Maximum (a) laser amplitude  and (b) dephasing strength as function of changes in the temperature $|\lambda_f | =\hbar \nu  \beta_f $ for heating (orange background) and cooling (blue background) processes. Results are shown for states with constant squeezing amplitude, starting with $|\lambda_i|=1$ and $\phi_{i}=\phi_{f}=0$.  Plots are for $t_f=1$. \label{fig:temp}}
\end{figure}

\begin{figure}
\centering 
\includegraphics[width = 1\columnwidth]{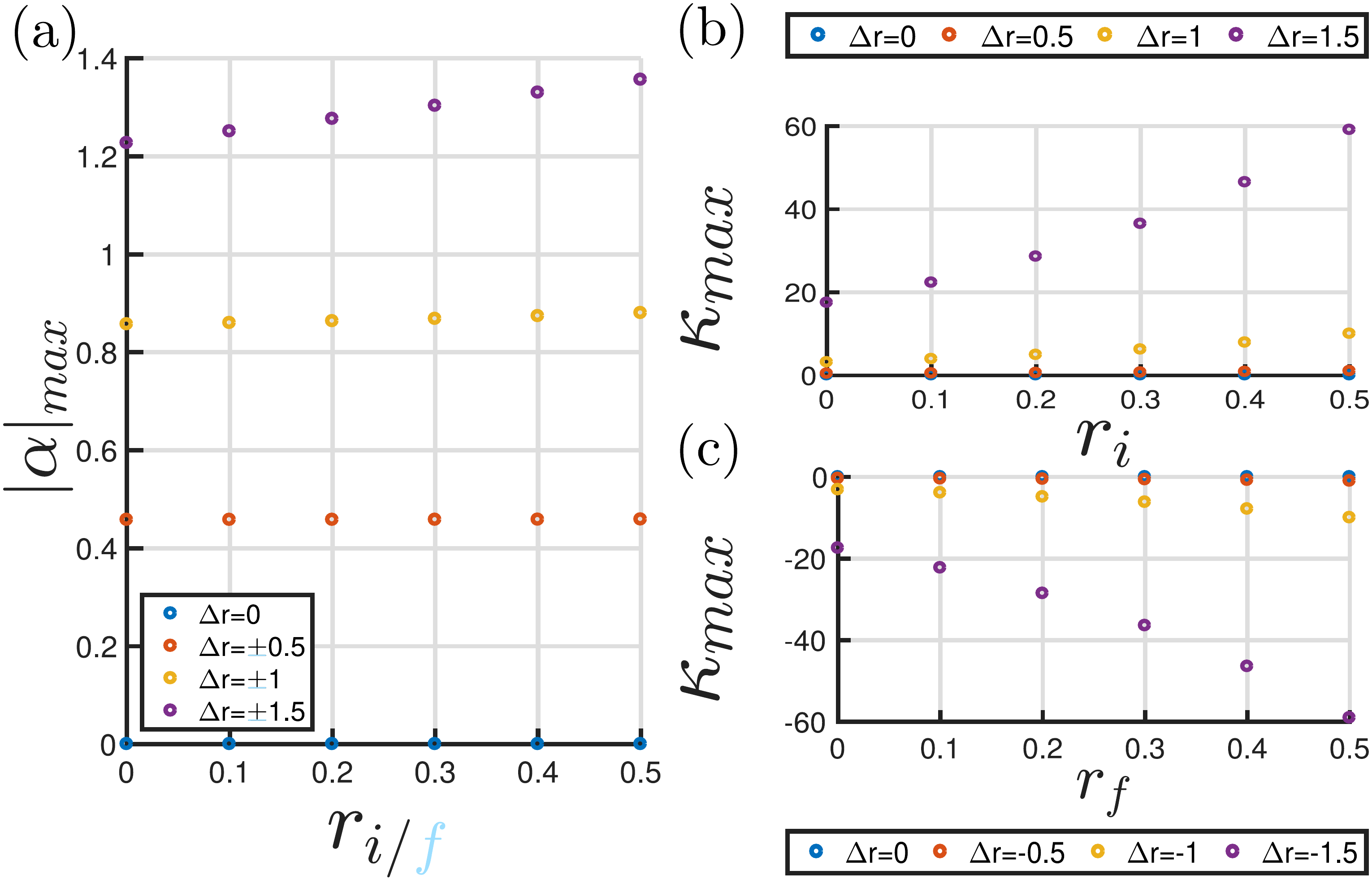}
\captionof{figure}{{\bf Maxima of the control parameters $\kappa_{\rm max}$ and $|\alpha|_{max}$} as function of the initial or final squeezing amplitudes, for different variation $\Delta r = r_f - r_i$.  
The two control parameters are `symmetric in squeezing', i.e. their maxima only depend on the absolute value $|\Delta r|$ of squeezing variation.   In other words,  a unique value of $|\kappa_{max}|$ is associated to a given couple of values ($r_{i},r_{f}$). 
Note that, while a high variation of the squeezing parameter is hard to engineer, only small values are needed since the variance exponentially depends on the squeezing amplitude---Eq. (\ref{eq:variance}). For instance $\Delta r=2$ drastically  reduces the variance by seven times.  Plots are for $t_f=1$. 
\label{fig:last}}
\end{figure}

Figure \ref{fig:control4laser} presents the control parameters for implementation of the dynamics for cooling, isothermal and heating processes. Interestingly, in the case of simple cooling and heating (with no squeezing),  the squeezing hamiltonian is not zero anymore, which is different from the former setup (cf. Fig. \ref{fig:control2laser}). Adding squeezing (dashed curves) leads to similar results. 
In turn, the parameter controlling the dephasing, $\kappa_t$, is   positive for heating and a negative for cooling, which matches with intuition. 
The influence of temperature and squeezing variations on the maxima of control parameters are presented in Figures \ref{fig:temp} and \ref{fig:last}.

 \smallskip

 \twocolumn

\clearpage

\onecolumn


%

\end{document}